\pdfoutput=1
\documentclass[useAMS,usenatbib]{mn2e}

\usepackage{graphicx}

\def\450um{450~$\mu$m}
\def\850um{850~$\mu$m}
\def\um{~$\mu$m}
\def\Spitzer{{\sl Spitzer}}

\title[A strongly starforming group]{A strongly starforming group: three massive galaxies associated with a QSO}
\author[F.~J. Carrera et al.]{F.~J. Carrera,$^{1,2}$\thanks{E-mail:
carreraf@ifca.unican.es} M.~J. Page,$^3$ J.~A. Stevens$^4$, R.~J. Ivison$^5$,
\newauthor 
T. Dwelly$^6$, J. Ebrero$^{7,1}$ and S. Falocco$^1$\\
$^{1}$Instituto de F\'\i{}sica de Cantabria (CSIC-Universidad de Cantabria), 39005 Santander, Spain\\
$^{2}$Astrophysics Group, Blackett Laboratory, Imperial College of Science, Technology and Medicine, Prince Consort Road, London SW7~2AZ\\
$^3$ Mullard Space Science Laboratory, University College London, Holmbury St. Mary, Dorking, Surrey RH5~6NT\\
$^4$ Centre for Astrophysics Research, University of Hertfordshire, College Lane, Hatfield, Herts AL10~9AB\\
$^5$ Astronomy Technology Centre, Royal Observatory, Blackford Hill, Edinburgh EH9~3HJ\\
$^6$ School of Physics and Astronomy, University of Southampton, Southampton, Hampshire SO17~1BJ\\
$^7$ SRON Netherlands Institute for Space Research, Sorbonnelaan 2, 3584 CA, Utrecht, The Netherlands\\
}

\begin{document}

\date{Accepted ???. Received ???; in original form 2010 July 27}

\pagerange{\pageref{firstpage}--\pageref{lastpage}} \pubyear{2010}

\maketitle

\label{firstpage}

\begin{abstract}

We present here photometric redshift confirmation of the presence of
large scale structure around the $z=1.82$ QSO RX~J0941, which shows an
overdensity of submm sources. Radio imaging confirms the presence of
the submm sources and pinpoints their likely optical-NIR
counterparts. Four of the five submm sources present in this field
(including the QSO) have counterparts with redshifts compatible with
$z=1.82$. We show that our photometric redshifts are robust against
the use of different spectral templates. We have measured the galaxy
stellar mass of the submm galaxies from their rest-frame $K$-band
luminosity obtaining $\log(M_*/M_\odot)\sim11.5\pm0.2$, slightly larger
than the Schechter mass of present day galaxies, and hence indicating
that most of the stellar mass is already formed. We present
optical-to-radio spectral energy distributions (SEDs) of the five SCUBA
sources. The emission of RX~J0941 is dominated by reprocessed AGN
emission in the observed MIR range, while the starburst contribution
completely dominates in the submm range. The SEDs of the other three
counterparts are compatible with a dominant starburst contribution
above $\sim24$~\um, with star formation rates SFR$\sim2000M_\odot$/yr,
central dust masses $\log(M_\mathrm{ dust}/M_\odot)\sim9\pm0.5$ and
hence central gas masses $\log(M_\mathrm{ gas}/M_\odot)\sim10.7$.
There is very little room for an AGN contribution. From X-ray upper
limits and the observed 24\um{} flux, we derive a maximum 2-10keV
X-ray luminosity of $10^{44}$~erg/s for any putative AGN, even if they
are heavily obscured. This in turn points to relatively small black
holes with $\log(M_\bullet/M_\odot)\la 8$ and hence stellar-to-black
hole mass ratios about one order of magnitude higher than those
observed in the present Universe: most of their central black hole
masses are still to be accreted. Local stellar-to-black hole masses
ratios can be reached if $\sim1.3$\% of the available nuclear gas mass
is accreted.  \end{abstract}


\section{Introduction}

It is currently commonly accepted that structure in the Universe is
formed through hierarchical processes, in which smaller structures
collapse to form galaxies, which then group together or coalesce to
form larger galaxies and galaxy groups and clusters. Within this
general background, ``anti-hierarchical" behaviour has been observed,
in the sense of the most dense large-galaxy-size structures collapsing
earlier and evolving faster than smaller galaxies \citep[e.g.][and
references therein]{Renzini06}. These processes are thought to be
accompanied by channeling of material to the central regions, where it
form stars and feeds a growing black hole  (BH). The feedback of
the latter regulates galaxy formation in an ``evolutionary sequence"
\citep[e.g.][]{SR98,Fabian99,Granato04,P04,King05,DM05,S05}.
According to these models galaxies grow through star formation and at
first host ``small" BH possibly in very dense and obscured nuclear
environments. As time progresses the combined radiative and mechanical
output from star formation and the BH accretion grows too, until the
AGN reaches QSO luminosities and literally blows away the
circumnuclear material.  This effectively terminates star formation
and the SMBH shines briefly (in cosmic terms) as an unobscured
QSO. Once it has accreted the remaining material, the QSO switches
off, leaving a passively evolving galaxy with a ``dormant" SMBH, such
as those observed in present-day massive galaxies \citep{MH04}.

The Far-infrared (FIR)-submillimetric (submm) spectral region is
ideally suited for searching for star formation, since the
black-body-like (greybody) emission by cold dust associated with star
formation peaks at those wavelengths. At the same time it suffers very
modest obscuration from the circumnuclear material.  This is proved by
the large numbers of strongly star forming submm galaxies (SMGs)
detected by SCUBA, AzTEC, SMA and, lately, Herschel, among other
facilities.

There are many multi-wavelength studies of SMGs, both from
surveys in "blank fields"
\citep[e.g.][]{SIB97,Hughes98,Eales99,Scott02,B05,Alexander05a,Coppin06,Laird10}
and targeting overdensities of such sources around high redshift
objects (e.g.,
\citealt{Kurk00,Pente00,Ivison00,S03,Smail03,DB04,Greve04,S04,Venemans07,Priddey08,Chapman09}).
These studies probe the properties of the star formation and of the
AGN, and the association between both phenomena, although there are
not many cases in which this association has been proved. SMGs have
substantial stellar masses in place by $z\sim2$. A relatively high
fraction of SMGs host AGN (e.g. \citealt{Alexander05a}, but see
\citealt{Laird10}), which seem to have BH-to-stellar mass ratios
smaller than local galaxies \citep{B05,Alexander08}. This implies that
BH growth lags galaxy growth, since the host galaxies are already
mature, while the BHs still require substantial growth (e.g., by about
a factor of 6), which however can be accommodated within the limits
imposed by the lifetime of the submm-bright phase \citep{Alexander08},
if accretion occurs close to the Eddington limit
\citep{Eddington13,Rees84}. The observed population of $z\sim2$ SMGs
would be sufficient to account for the formation of the population of
bright ellipticals seen in the present universe \citep{Swinbank06}.

We have found \citep{P04,S05} a sample of X-ray-obscured QSOs at
$z\sim 2$ (when most of the star formation and BH growth is occuring
in the Universe) with strong submm emission, much higher than
X-ray-unobscured QSOs at similar redshifs and luminosities,
which however represent 85-90\% of the X-ray QSOs at that epoch. This
submm emission is too high to come from the AGN \citep{P01,S05}. The
fields around these objects show strong overdensities of SMGs
\citep{S04,S10}. The UV and X-ray spectra of the central QSOs
\citep{P10} show evidence for strong ionized winds which produce the
X-ray obscuration. Piecing all these clues together we infer that the
host galaxies of these QSOs are undergoing strong SFR, while the
central SMBH are also growing through accretion. The ionized winds are
strong enough to quench the star formation, so the QSOs must be just
emerging from a strongly obscured accretion state to become ``normal"
unobscured QSOs with passively evolving galaxies, in an evolutionary
stage which might last about 10-15\% of the QSO lifetime. They appear
to be in the centres of high density peaks of the Universe.

In this paper we endeavour to prove the physical association of one of
those overdensities to its central QSO (RX~J094144.51+385434.8,
henceforth RX~J0941). We also investigate the nature and evolutionary
stage of those SMGs through their rest frame UV-FIR SEDs.

The outline of this paper is as follows: in Section \ref{data} we
summarize the data used in this paper and the reduction process, in
Section \ref{zphot} we describe how the source catalogues in each band
have been merged into a single catalogue, and how that catalogue has
been used to obtain photometric redshifts. These are then used in
Section \ref{Results} to show which objects are associated with the
structure around the QSO and to the submm sources, and how we have
obtained the different physical parameters for each object. The
evolutionary stage of our sources is discussed in Section
\ref{Discussion}. Finally, in Section \ref{Conclusions} we summarize
our results. We have assumed throughout this paper a Hubble constant
$H_0=70$~km/s/Mpc, and density parameters $\Omega_\mathrm{ m}=0.3$ and
$\Omega_\Lambda=0.7$. The spectral index $\alpha$ is defined as
$F_\nu\propto \nu^\alpha$.

\begin{figure*}
  \includegraphics[width=18cm]{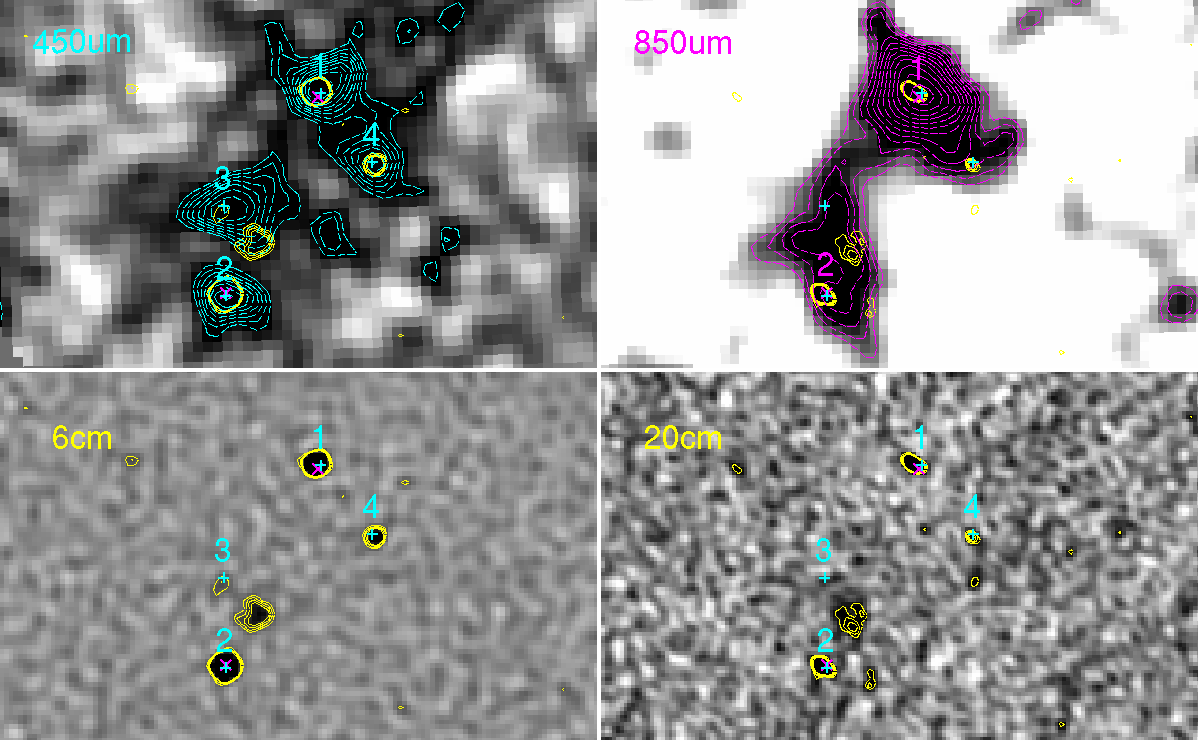}

  \caption{Finding charts of a $2.1'\times1.3'$ region around the submm
  structure associated with RX~J0941 (N is up and E to the left).
  The region around 850\_3 (to the NW of RX~J0941) is shown
  sepparately in Fig. \ref{Figfc850-3}.
  Top left: 450\um{} SNR image with 2 to 6$\sigma$ contours (in steps
  of 0.5$\sigma$, dashed cyan) and 6cm 2 to 5$\sigma$ contours (in steps of
  1$\sigma$, yellow), the numbers and crosses in cyan correspond to
  the 450\um{} sources in Table~\ref{TabSCUBA}. Top right: 850\um{} SNR
  image with 2 to 8$\sigma$ contours (in steps of 0.5$\sigma$,
  dashed magenta) and 20cm 2 to 5$\sigma$ contours (in steps of 1$\sigma$,
  yellow), the numbers and ``x" in magenta correspond to the 850\um{}
  sources in Table~\protect{\ref{TabSCUBA}} (the cyan crosses to the 450\um{}
  sources in that Table). Bottom left: 6cm image and contours (as
  above) with source numbers and crosses as above (note that
  sources 1 and 2 are common to 450\um{} and 850\um). Bottom right: 20cm
  image and contours (as above) with source numbers and markers as
  above. In greyscale, magenta appears slightly darker than cyan
  and the steeper yellow contours appear white.} \label{FigSMMradio}

\end{figure*}

\begin{table*}
 \centering
 \begin{minipage}{160mm}
  \caption{SCUBA source positions and fluxes (see Section
  \ref{data}). The first part of the table corresponds to
  850\um{} and the second to 450\um. Our source number for the
  counterpart is given under \# (see Table~\ref{TablePhot} and Section~\ref{SEDs}), as well
  as the best fit photometric redshift ($z_p$) and its $1\sigma$ confidence interval
  interval. Star formation rates (SFR) and
  8-1000$\mu$m IR luminosities are from template SEDs (see
  Section~\ref{SEDs}). Dust mass ranges and galaxy
  stellar masses are derived in Sections~\ref{Mdust} and \ref{Mstar}
  respectively. Infrared luminosities, SFR, dust masses and galaxy
  stellar masses are obtained assuming $z=1.82$}
  \begin{tabular}{lccccccrrrr}
  \hline
Source & R.A.        & Dec.               & Flux          &  \# & $z_p$ & $1\sigma$ & $L_{8-1000\mu\mathrm{m}}$ & SFR    & $M_\mathrm{dust}$ & $\log M_*$$^b$ \\ 
       & (h m s)     & ($^\circ\,'\,''$) & (mJy)          &     &       &           & ($10^{13}L_\odot$) & ($M_\odot$/yr) & ($10^8 M_\odot$)   & ($\log M_\odot$)\\
 \hline							            
850\_1$^1$ & 09 41 44.63 & +38 54 38.88   & 12.4$\pm$1.1  &  39 &  -    &    -      &  3.9   & 4800$^a$ 
     &   8-25 & -    \\
850\_2$^2$ & 09 41 46.31 & +38 53 56.94   &  5.7$\pm$2.1  & 140 & 1.85  & 1.83-2.16 &   1.3   & 1700     
     &   3-12 & 11.6 \\
850\_3     & 09 41 42.62 & +38 55 36.81   &  3.1$\pm$2.2  & 501 & 1.85  & 1.83-2.20 &   0.5   &  650     
     &    2-6 & 11.5 \\ \hline						
	                  	       	           
450\_1$^1$ & 09 41 44.57 & +38 54 39.46   & 44.7$\pm$10.2 &  39 &  -    &    -      &   2.4   & 2900     
     &   5-22 & -    \\
450\_2$^2$ & 09 41 46.31 & +38 53 56.22   & 41.2$\pm$10.0 & 140 & 1.85  & 1.83-2.16 &   2.2   & 2700     
     &   4-23 & 11.6 \\
450\_3     & 09 41 46.35 & +38 54 15.44   & 33.5$\pm$8.8  & 262 & 2.78  & 2.58-3.00 &   1.8   & 2200     
     &   4-16$^c$ & 10.9$^c$ \\
450\_4     & 09 41 43.63 & +38 54 24.77   & 28.8$\pm$7.8  &  98 & 1.84  & 1.81-2.08 &   1.4   & 1700     
     &   3-16 & 11.5 \\
 \hline
\end{tabular}

\noindent $^1$ 850\_1 and 450\_1 are the same physical source: the central QSO.\
\noindent $^2$ 850\_2 and 450\_2 are the same physical source\\
\noindent $^a$ Above maximum $L_\mathrm{ IR}$ luminosity for \citet{CE01}\\
\noindent $^b$ Stellar masses are for the counterparts under column ``\#''\\
\noindent $^c$ The best fit redshift is 2.78, very different from 1.82. For the 
higher redshift SFR=2600$M_\odot/$y, $L_\mathrm{ 8-1000\mu m}=2.1\times 10^{13}L_\odot$,
$\log(M_*/M_\odot)=11.3$. Using the $1\sigma$ uncertainty in the photometric redshift the uncertainty interval becomes $M_\mathrm{ dust}=(4-25)\times10
^8 M_\odot$\\
\end{minipage}
\label{TabSCUBA}
\end{table*}

\begin{table*}
\centering
 \begin{minipage}{140mm}
  \caption{Summary of the observations on the field around RX~J0941}
  \begin{tabular}{llrlrcrr}
  \hline
Telescope & Instrument & Date & Filter/channel      & Texp & $m_0$  & $1\sigma$ detection level & Seeing \\
          &            &      &                     & (s)  & (Vega) &                          & (FWHM, $''$)\\ 
\hline
WHT          & PFI    & 28-May-2003 & R Harris      &  2880 & 32.35 & 25.40$^1$ & 0.993 \\
Gemini-North & GMOS   & 13-Feb-2005 & i             &  3000 & 34.34 & 25.85$^1$ & 0.743 \\
INT          & WFC    & 03-Mar-2006 & Z             & 72000 & 29.49 & 24.85$^1$ & 1.275 \\
UKIRT        & UFTI   & 11-Feb-2004 & J             &  7200 & 30.18 & 23.02$^1$ & 0.741 \\
UKIRT        & UFTI   & 24-May-2003 & K             &  7700 & 28.90 & 21.87$^1$ & 0.708 \\
\hline
\Spitzer{}      & IRAC   & 06-May-2005 & 4.5~$\mu$m &  3000 &   -   &   0.5$^2$ & - \\ 
\Spitzer{}      & IRAC   & 06-May-2005 &   8~$\mu$m &  3000 &   -   &   2.1$^2$ & - \\ 
\Spitzer{}      & MIPS   & 12-Apr-2005 &  24~$\mu$m &   336 &   -   &    52$^2$ & - \\ 
\hline
VLA          & C band & 25-Mar-2008 & 4860~MHz      & 64000 &   -   &   6.6$^3$ & -$^4$ \\ 
GMRT         & L band & 21-Jan-2008 & 1280~MHz      & 25000 &   -   &    29$^3$ & -$^5$ \\
\hline
\end{tabular}

\noindent $^ 1$: in Vega magnitudes, from App. \ref{AppMatchPhot} using 1.5~arcsec aperture and the ``standard" aperture correction\\
\noindent $^ 2$: in $\mu$Jy, see App. \ref{AppMatchPhot}\\
\noindent $^ 3$: in $\mu$Jy, see Sec. \ref{data}\\
\noindent $^ 4$: beam size $3.3''\times2.2''$, position angle 57~deg\\
\noindent $^ 5$: beam size $3.6''\times3.5''$, position angle 73~deg\\
\end{minipage}
\label{TableData}
\end{table*}

\section[]{Data}
\label{data}

The $R$-band, $K$-band, \Spitzer{} and SCUBA data used here are the
same ones discussed in \citet{S04,S10} (henceforth S10), but with a
new reduction of the $R$-band data, all the astrometry tied to this
band (see below) and fluxes for fainter \Spitzer{} sources. The $i$, $Z$
and $J$-bands and radio data are newly presented here.  We show in
Fig.~\ref{FigSMMradio} the central area of the submm images along with
contours and images of the radio data. The region to the NW around
source 850\_3 is shown in Fig.~\ref{Figfc850-3}.

We give in Table~\ref{TabSCUBA} the deboosted 850\um{} and 450\um{}
fluxes from S10, where we have symmetrized the error bars taking the
worst of the upper and lower error bars. The source extraction method
used in S10 \citep[described in][]{Scott02} can miss real sources
separated by less than a beam (14.2$''$ for 850\um). This is probably
the case for 450\_3, which is blended together with 850\_2 (see
Fig.~\ref{FigSMMradio} and S10): in 850~\um{} these two sources are
part of a $\sim$N-S struture with 850\_2/450\_2 in the Southern tip
and 450\_3 close to the Northern tip.

A summary of the origin of the observed optical-to-radio data used
here is included in Table \ref{TableData}, as well as some of the
characteristics of the final images. The data come from several
different telescopes and observatories. At all wavelengths shorter
than radio, the data were taken as individual images which were later
combined using standard procedures (whose details may differ from data
set to data set and are outlined below). Since this involved
combination of images, often with non-integer pixel shifts and
individual re-scaling, we have used a simulation technique to take
into account the effect of this procedure on the overall photometric
calibration (see Section \ref{photcal}).

The astrometry of the $R$-band image was matched to
APM\footnote{{\tt http://www.ast.cam.ac.uk/$\sim$apmcat}}. The
astrometry of the $iZJK$ and \Spitzer{} images was then refered to the
$R$-band image. The astrometry of the SCUBA images in S10 was also tied
to the $R$-band image, allowing for a shift, asymmetric X and Y scales
and rotation with respect to the ``native'' SCUBA astrometric
system. The sources used as reference for these were 850\_1, 850\_2
and 850\_3 for the 850~$\mu$m image and 450\_1, 450\_2 and 450\_4 for
the 450~$\mu$m image. 450\_3 was excluded because it is probably a
blend of several sources (see above).

The radio data have been reduced with {\sc AIPS} following a standard
procedure. For the VLA data we have used an intermediate weighting
scheme between natural and uniform ({\tt ROBUST=0} in {\sc
AIPS}). This gives a synthesised beam of $3.7''\times3.5''$ and a
noise of 6.6~$\mu$Jy/beam. We have used the original astrometry for
the radio images. The very good angular resolution and
astrometry of the radio images allow relating better the submm
detections to the corresponding optical counterparts.

Finally, in Section~\ref{AGNX-ray} we have used the upper limit server
{\sc FLIX}\footnote{\tt http://ledas-www.star.le.ac.uk/flix/flix.html}
(which in turn uses the procedure outlined in \citealt{C07}) to obtain
$1\sigma$ upper limits to the 0.5-2~keV (soft) and 2-12~keV (hard)
X-ray countrates from the corresponding background maps on the
positions of the SCUBA sources. These X-ray data are the same ones
discussed in \citet{P10}, but the server uses the standard output
from the {\sl XMM-Newton} pipeline.

\begin{table*}
\centering
 \begin{minipage}{180mm}
  \caption{Positions and photometry (Vega magnitudes and fluxes) of the counterparts mentioned in the text}
  \begin{tabular}{rcccccccccc}
  \hline
    \#        & R.A.        &  Dec.               & $R$ & $i$ & $Z$ & $J$ & $K$ & $F_{4\mu{\mathrm m}}$ & $F_{8\mu{\mathrm m}}$ & $F_{24\mu{\mathrm m}}$ \\
              & (h m s)     & ($^\circ\,'\,''$)   &     &     &     &     &     & ($\mu$Jy)  & ($\mu$Jy)  & ($\mu$Jy) \\
 \hline
 13 & 09:41:45.84 & +38:53:53.54 & 22.27$\pm$0.10 & 21.42$\pm$0.07 & 20.95$\pm$0.04 & 19.87$\pm$0.18 & 18.12$\pm$0.11 & 37$\pm$2 & 15$\pm$2 & - \\ 
 16 & 09:41:46.14 & +38:53:55.71 & 23.36$\pm$0.17 & 22.76$\pm$0.09 & 22.82$\pm$0.17 & 21.12$\pm$0.25 & 19.07$\pm$0.13 & - & - & - \\ 
 29 & 09:41:46.17 & +38:54:17.99 & 23.27$\pm$0.18 & 22.84$\pm$0.10 & 22.82$\pm$0.18 & 21.50$\pm$0.33 & 19.95$\pm$0.23 & 13$\pm$1 & - & - \\ 
 39 & 09:41:44.64 & +38:54:39.57 & 19.63$\pm$0.06 & 18.73$\pm$0.07 & 18.79$\pm$0.02 & 17.70$\pm$0.17 & 16.46$\pm$0.10 & 715$\pm$36 & 1661$\pm$87 & 5443$\pm$360 \\ 
 98 & 09:41:43.56 & +38:54:24.33 & 23.70$\pm$0.51 & 23.00$\pm$0.18 & 22.20$\pm$0.21 & 20.59$\pm$0.30 & 18.26$\pm$0.13 & 58$\pm$3 & 37$\pm$3 & - \\ 
 135 & 09:41:45.94 & +38:54:14.66 & 24.56$\pm$0.52 & 23.32$\pm$0.13 & 22.84$\pm$0.17 & 21.41$\pm$0.30 & 19.51$\pm$0.16 & 19$\pm$1 & 13$\pm$2 & - \\ 
 140 & 09:41:46.31 & +38:53:56.55 & 23.21$\pm$0.23 & 22.70$\pm$0.12 & 21.96$\pm$0.12 & 20.07$\pm$0.20 & 18.03$\pm$0.11 & 85$\pm$4 & 54$\pm$3 & 699$\pm$69 \\ 
 262 & 09:41:46.35 & +38:54:13.94 & 25.26$\pm$0.96 & 25.88$\pm$1.09 & - & 26.80$\pm$34.38 & 21.13$\pm$0.54 & 14$\pm$1 & 14$\pm$2 & - \\ 
 501 & 09:41:42.65 & +38:55:36.37 & 22.95$\pm$0.09 & 22.86$\pm$0.08 & 22.08$\pm$0.07 & - & - & 69$\pm$4 & 51$\pm$3 & 345$\pm$57 \\ 
 512 & 09:41:42.53 & +38:55:36.85 & 23.84$\pm$0.16 & 22.67$\pm$0.08 & 22.82$\pm$0.12 & - & - & - & - & - \\ 
  \hline
\end{tabular}
\end{minipage}
\label{TablePhot}
\end{table*}

\subsection{Photometric calibration}
\label{photcal}

We took advantage of the existence of SDSS \footnote{{\tt
http://www.sdss.org/}} \citep{SDSSDR5} data in the RX~J0941 field to
tie our optical photometry to the SDSS DR5 photometric calibration.
In summary, we obtained Vega magnitudes for the SDSS sources within
the FOV of each of our images \citep[using expressions from][and from
the SDSS and WFC survey web pages]{Jester05}, using only point-like
objects (type=6 in the SDSS source classification). We then matched
the SDSS sources to the sources in our images, fitting an additive
constant between the {\tt magbest} {\sc SExtractor} magnitudes of our
sources (excluding saturated ones) and the Vega magnitudes, taking
into account the statistical errors of both sets of magnitudes. These
additive constants are the zero points given in Table
\ref{TableData}. The error in the zero point is added in quadrature to
the error in the magnitude of each source to get the final error in
the magnitude of each source.

Unfortunately, the FOV of the $JK$ images was very small and there
were no 2MASS sources within them to provide an independent check on
the magnitudes. However, the sky conditions while our NIR images were
taken were photometric and the templates fits (see Section
\ref{zphot}) do not show any systematic trend for the NIR magnitudes.

The \Spitzer{} source lists in S10 had a high significance
detection threshold, tailored to find counterparts to the submm
sources. However, the deep $RiJK$ images used here detected much
fainter sources, which could be seen in the \Spitzer{} images
but were not present in those source lists. In order to have source
lists which would include those fainter sources and to keep
compatibility with the results of S10 we used {\sc SExtractor}
with lower significance to get aperture magnitudes (1.3, 1.7 and
3.2~arcsec radius for 4.5, 8 and 24~$\mu$m respectively, chosen to
avoid source confusion) for the sources, and then obtained an
empirical factor (taking into account the errors on both axis) which
would match these {\sc SExtractor} fluxes with those of S10
for the common sources. The {\sc SExtractor} fluxes were then
multiplied by these factors, taking also into account the
uncertainties in the factors to calculate the final flux uncertainties.
These final \Spitzer{} fluxes are hence corrected to ``infinite'' aperture.

\section[]{Photometric redshifts}
\label{zphot}

\begin{table*}
 \centering
 \begin{minipage}{140mm}
  \caption{Positions and fluxes for the radio sources from VLA (4.86~GHz) and GMRT (1.28~GHz), as well as the spectral index 
$\alpha$. The upper limits quoted are 3$\sigma$.}
  \begin{tabular}{lccccccc}
  \hline
Source        & R.A.        & $\Delta$R.A. & Dec.               & $\Delta$Dec. & 1.28~GHz flux & 4.86~GHz flux & $\alpha$ \\ 
              & (h m s)     & (s)          & ($^\circ\,'\,''$) & ($''$)       & ($\mu$Jy)     & ($\mu$Jy)     \\
 \hline
850\_1/450\_1 & 09 41 44.66 & 0.01 & +38 54 39.9 & 0.1 & 650$\pm$29  & 194$\pm$10   & -0.90$\pm$0.05 \\
850\_2/450\_2 & 09 41 46.32 & 0.01 & +38 53 56.7 & 0.1 & 674$\pm$52  & 313$\pm$11   & -0.57$^{+0.07}_{-0.06}$ \\ 
850\_3        & 09 41 42.64 & 0.03 & +38 55 36.8 & 0.3 & $<$96       & 28.0$\pm$6.3 & $>$-0.92 \\ 
450\_3        & 09 41 46.39 & 0.03 & +38 54 13.8 & 0.3 & $<$81       & 28.5$\pm$6.1 & $>$-0.78 \\ 
450\_4        & 09 41 43.59 & 0.01 & +38 54 24.3 & 0.1 & 189$\pm$29  & 71.8$\pm$6   & -0.72$^{+0.14}_{-0.12}$ \\ 
 \hline
Blob          & 09 41 45.76 & 0.03 & +38 54 07.8 & 0.3 & 800$\pm$100 & 120$\pm$20 & -1.42$^{+0.18}_{-0.14}$ \\
\end{tabular}
\end{minipage}
\label{TableRadio}
\end{table*}

Once we have the final images in each band, we have run {\sc
SExtractor} on each one of them with fairly low significance
requirements (3 or more pixels 3 or more $\sigma$ above the
background) to obtain comprehensive source lists in each band. These
source lists have been restricted to the FOV of the $JK$ images
(roughly 1.5~arcmin squared), to ensure the maximum wavelength
coverage for the photometric redshifts. If a source was detected in
either of the \Spitzer-IRAC bands, we have included the corresponding
measurements, otherwise, we have just used the 5 optical-NIR bands
$RiZJK$ to assign the photometric redshifts. The individual $RiZJK$
and IRAC source lists have been merged, considering two sources to be
the same if they are closer than 1~arcsec to each other. The
positions, magnitudes and fluxes of the counterparts mentioned below
are given in Table~\ref{TablePhot}.

We have a total of 239 unique sources detected in at least one of
our 8 bands. Photometric redshift fits could not be performed on 14 of
them, mostly because at least four of the five bands were outside the
FOV and, in two cases, because the two measurements which were within
the FOV were upper limits.

To obtain photometric redshifts we have used {\sc hyperz}
(version 11, \citealt{BMP00}). We have not allowed for intrinsic
reddening but we have included Galactic de-reddening
($E(B-V)=0.015$). We have allowed a redshift interval between 0 and
3. Many current photometric redshift codes \citep[e.g.][]{B00} take
into account the expected apparent magnitude distribution of the
objects (using galaxy luminosity functions), reducing the likelihood
of redshifts that would result in absolute magnitudes too bright or
too dim for the observed magnitudes. {\sc hyperz} does not provide
this facility, so we have instead checked ``a posteriori'' the
absolute $B$-band magnitude of the objects. Following \citet{MRR08},
henceforth MRR08, we have adopted redshift-dependent lower and upper
limits to the absolute blue magnitude of max(-19.5,-17-$z$) and
max(-25,-22.5-$z$), respectively.


We have tried several sets of templates \citep{BC03,MRR08} and
settled for the ``new" galaxy templates described in MRR08. There are
7 different templates (starburst SB, youngE, E, Sab, Sbc, Scd,
Sdm). The last 5 templates define a sequence of increasingly later
galaxies, and template youngE is a ``young" Elliptical with an age of
1Gyr, to allow for the limited time for evolution available at higher
redshifts. These templates do not include dust emission, so they are
appropriate until about 3\um{} rest-frame, including our observed
$RiZJK$ and IRAC bands at the relevant redshifts. When comparing the
full SED we will complement those with other SEDs with a wider
coverage (see Section \ref{SEDs}).
 
For a good match of the spectral shape of the templates to the
observed photometric points it is essential that photometry of each
source corresponds to the same aperture.
To minimise the mutual
photometric contamination by close sources in the optical-NIR bands,
we have used an adaptive aperture size with
a maximum radius of 1.5$''$ (see App.~\ref{AppMatchPhot}).

Out of the 225 sources in which photometric redshift fits could be
performed, if we select those with a 99\% redshift interval that
includes $z=1.82$ and that is narrower than 0.6 (full span from lower
to upper limits, i.e., relatively well constrained), we get 4 sources
including all the counterparts to the SCUBA sources (except for
450\_3).

Using the \citet{BC03} templates instead (which also include blue
templates with recent bursts of star formation) the best fit
photometric redshifts for the counterparts to 850\_2, 850\_3 and
450\_4 are the same as with the MRR08 templates, and these 3 sources
are also among the 8 with ``narrow" 99\% confidence interval including
$z=1.82$ with the \citet{BC03} templates.

We conclude that our photometric redshifts are robust against using
different sets of templates. In particular, the counterparts to the
SCUBA sources (see below) always appear amongst the sources with higher quality
fits compatible with the redshift of the central QSO.

\section[]{Results}
\label{Results}

\subsection{Radio data}
\label{RadioResults}

The list of the radio sources detected can be found in
Table~\ref{TableRadio}. The very good alignment of the radio sources
(with their original astrometry) to the submm sources (aligned
independently to the $R$-band image) can be seen in
Figs.~\ref{FigSMMradio} and \ref{Figfc850-3}. All submm sources are
detected with high SNR$> 4$ in the radio images (except for 450\_3,
which also has the optical-NIR counterpart less likely to be
associated with the QSO, see below). This increases our confidence both
in the existence of the submm sources and on their association with
particular optical-NIR sources, since the angular resolution of the
radio images allows very accurate positioning of the long wavelength
sources.

It is worth noting the existence of a ``blob'' of diffuse radio
emission to the SW of 450\_3 (09:41:45.694 +38:54:07.57), clearly
visible in both the VLA and GMRT data with a diameter of about
3~arcsec. It appears to be connected to 450\_3 by a bridge of $\sim
2\sigma$ emission in the former image. Its spectral index is
$\alpha=-1.65\pm 0.20$ which would advocate for it to be a lobe of
radio emission. This area of diffuse emission is also clearly devoid
of optical and \Spitzer{} counterparts, which would be
consistent with the lobe hypothesis. There is no evidence of a
counter-lobe to the NE of 450\_3. If it is a lobe, it could be
associated instead with source 850\_2/450\_2, which is much brighter
in radio, but no bridge of emission is seen between the latter source
and the ``blob''.

\begin{figure*}
\hbox{
  \includegraphics[width=9.5cm]{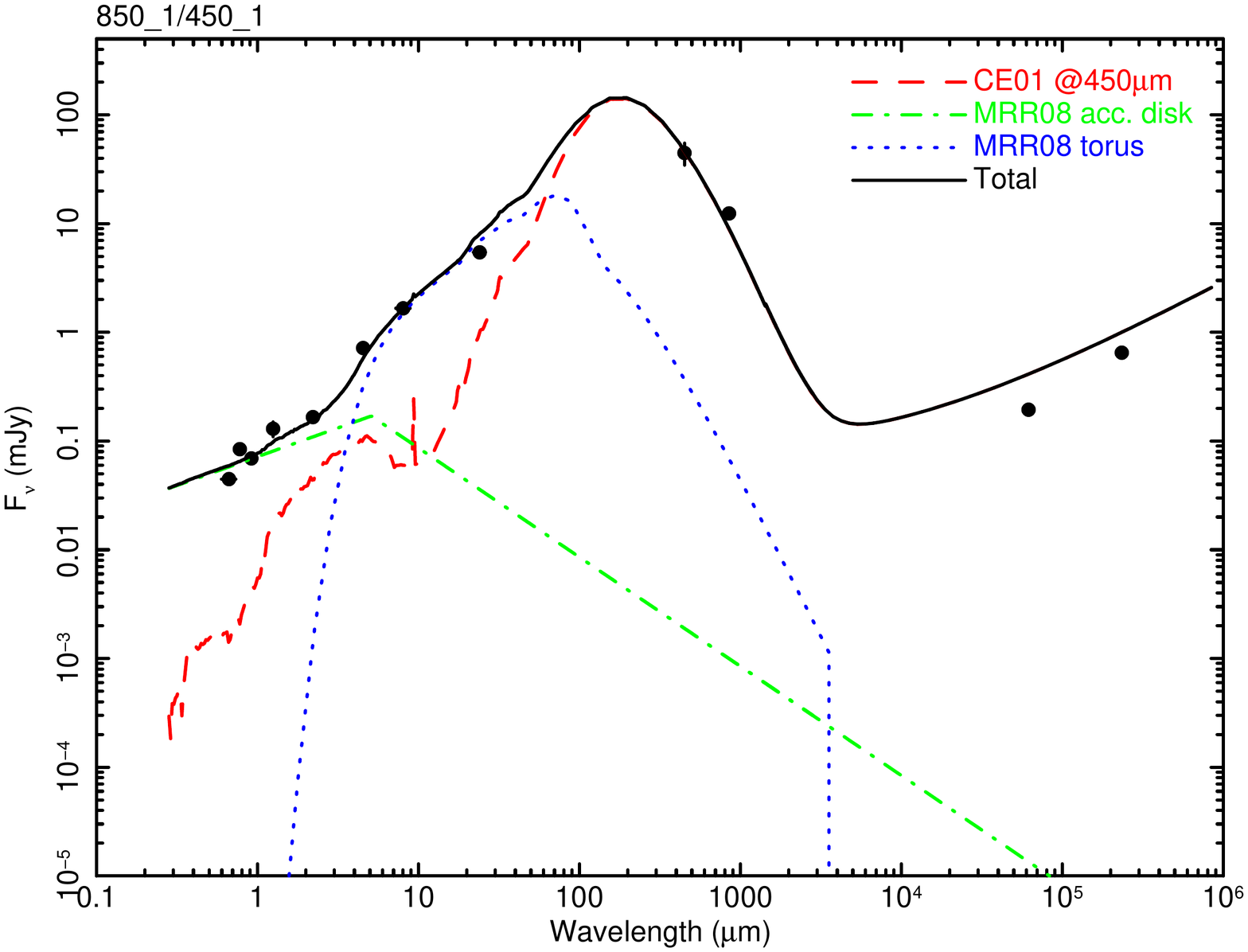}
\hspace{0.1cm}
  \includegraphics[width=9.5cm]{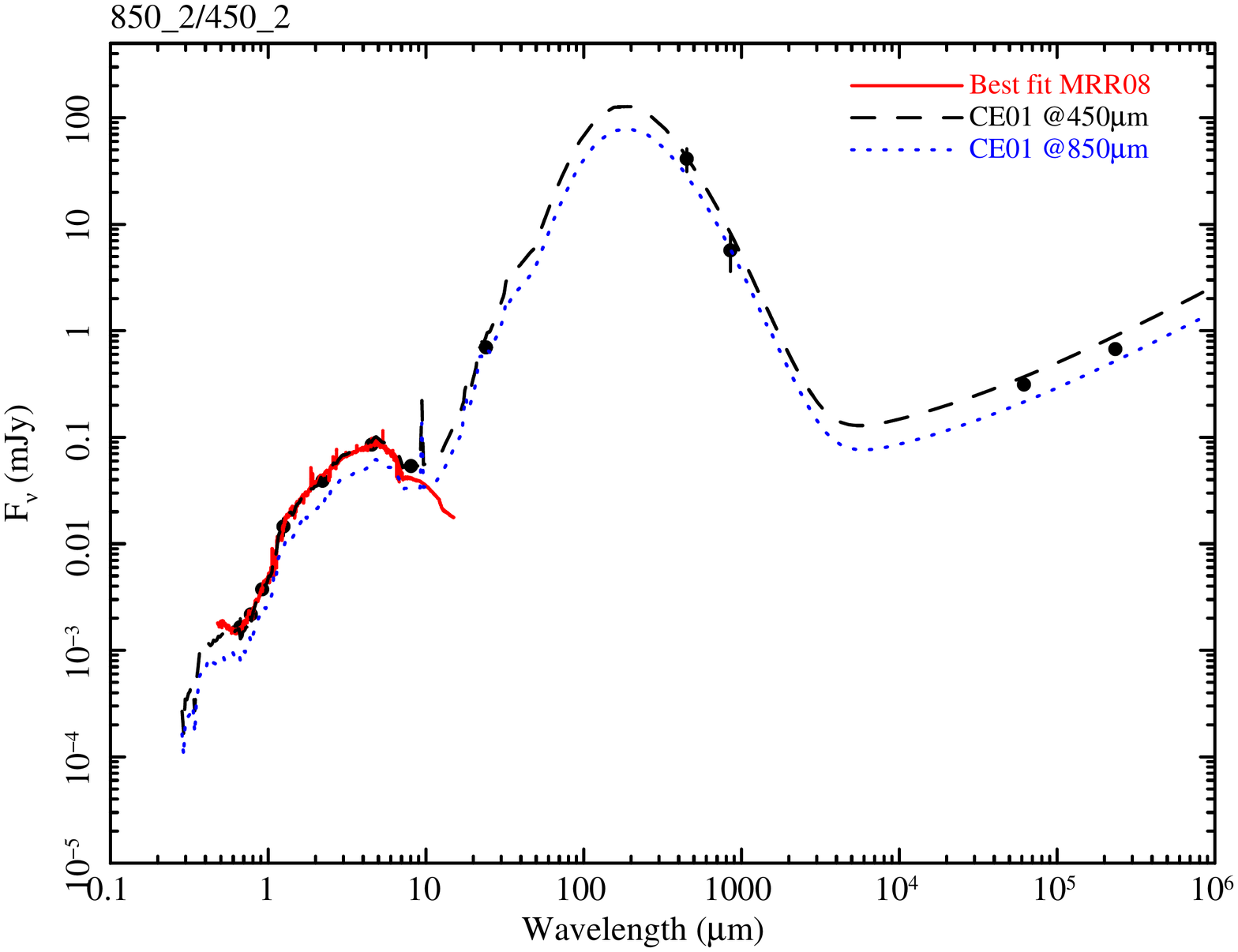}
}
\hbox{
  \includegraphics[width=9.5cm]{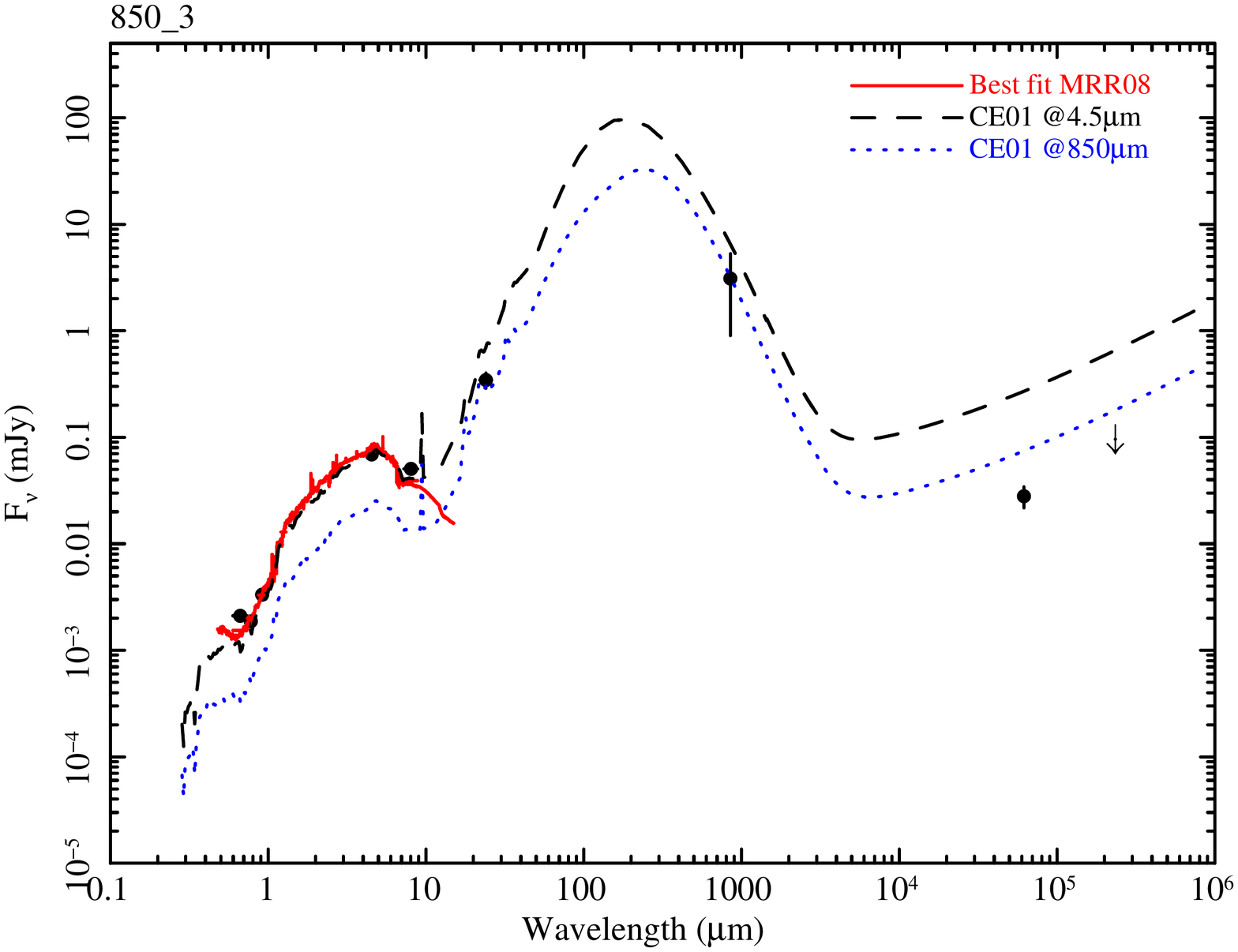}
\hspace{0.1cm}
  \includegraphics[width=9.5cm]{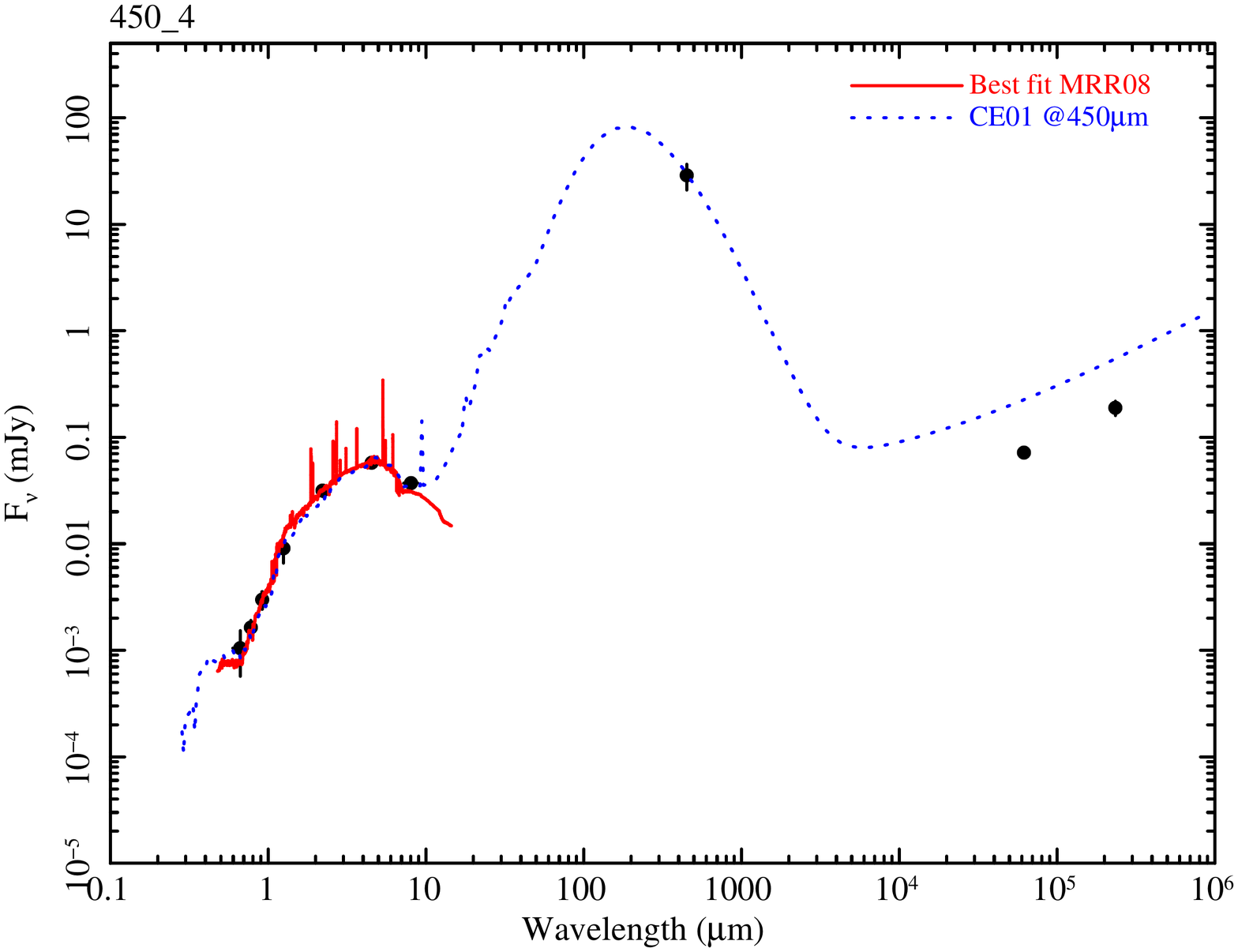}
}
\hbox {
  \includegraphics[width=9.5cm]{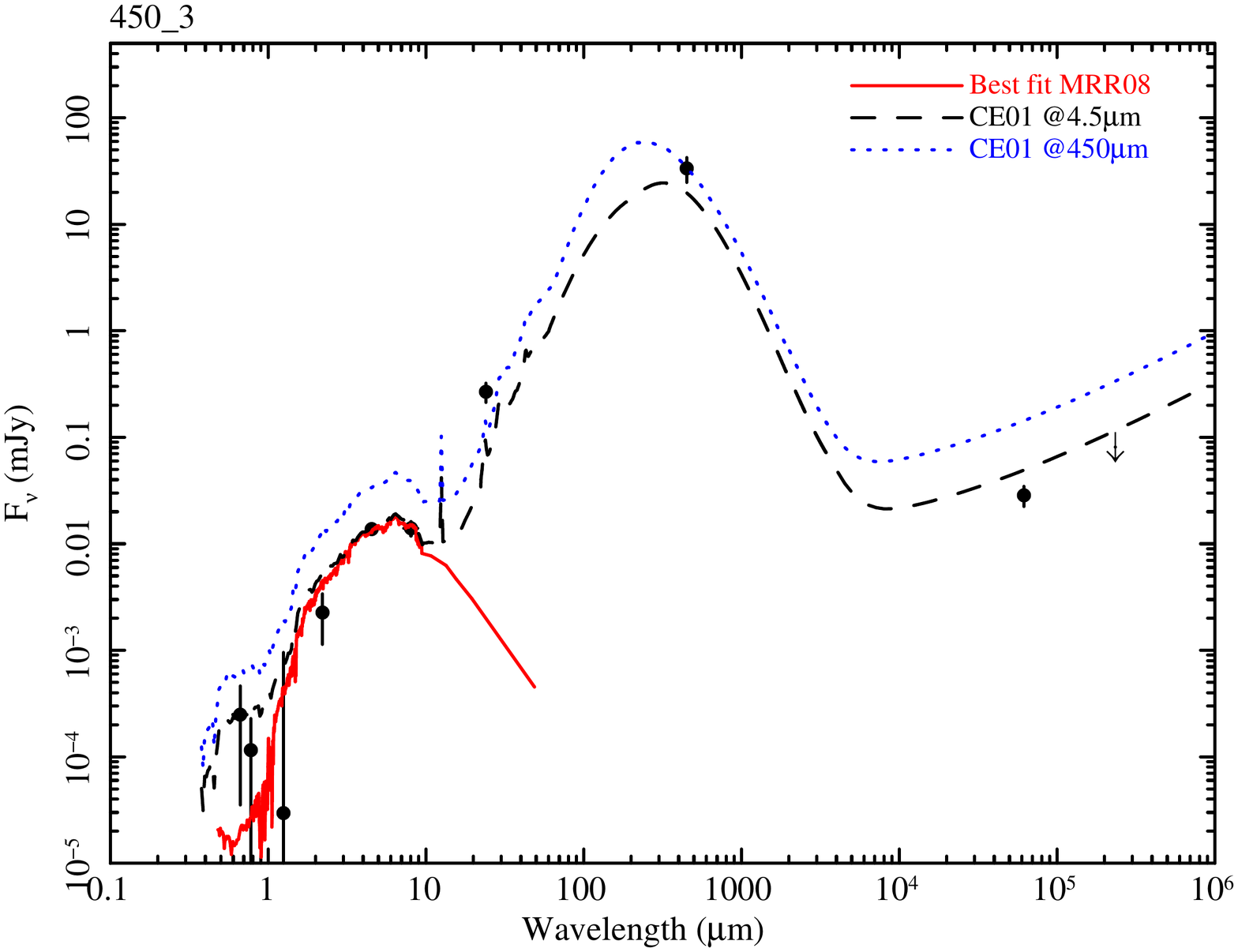}
\hspace{0.1cm}
  \includegraphics[width=9.5cm]{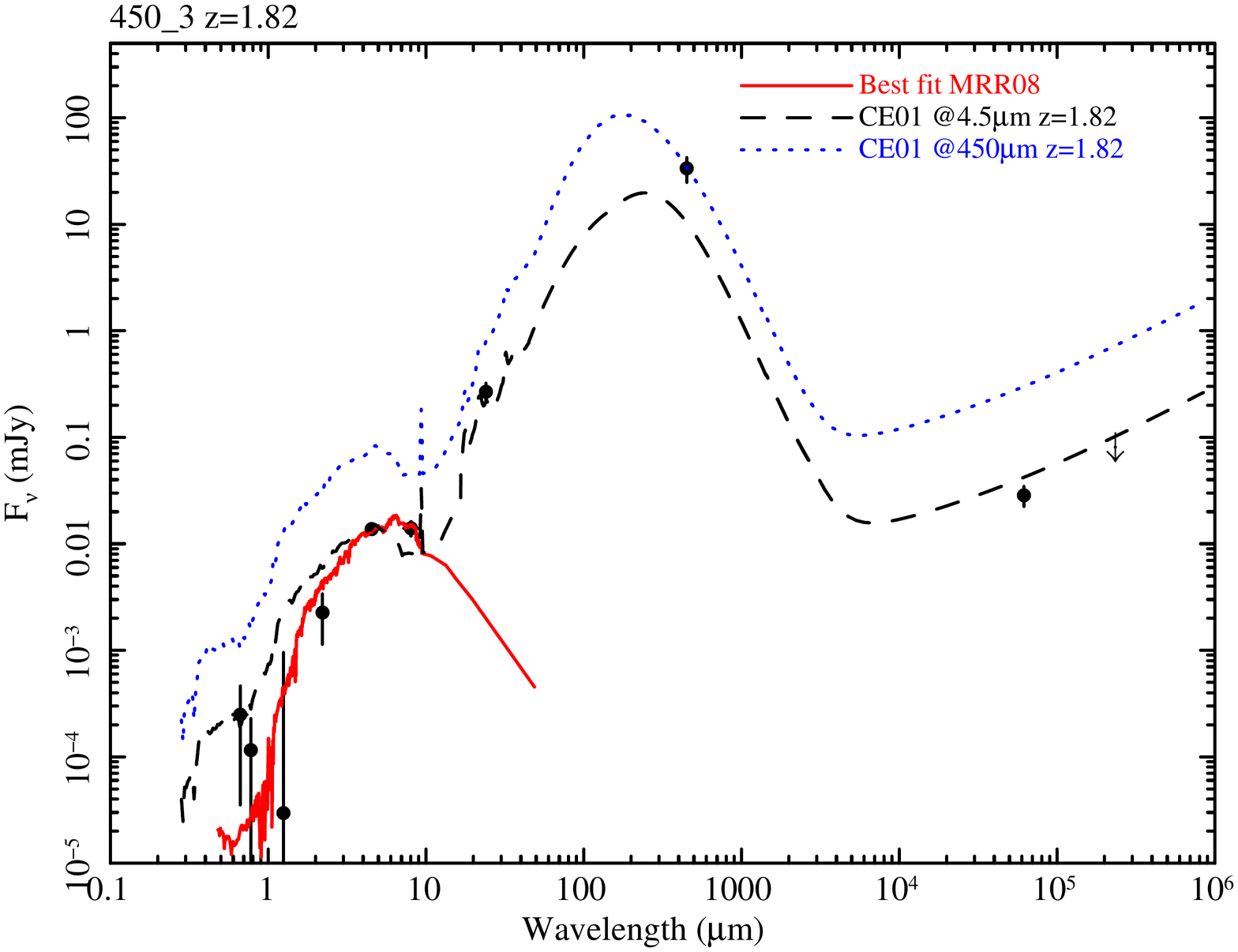}
}

  \caption{Spectral Energy Distribution in $F_\nu$  in the observer's frame for
  850\_1/450\_1 (RX~J0941, top left), 850\_2/450\_2 (top
  right), 850\_3 (middle left), 450\_4 (middle right),
  and 450\_3 (with templates at the best photometric redshift
  -bottom left- and fixed at $z=1.82$ -bottom right-). Observed points
  ($RiZJK$, 4.5, 8, 24, 450 and \850um, and 6 and 20 cm)
  are shown as black dots and upper limits as down-pointing
  arrows. The meaning of each line is summarized in each panel and
  explained fully in Section \ref{SEDs}.}  \label{FigSED}

\end{figure*}

\begin{figure*}
  \includegraphics[width=18cm]{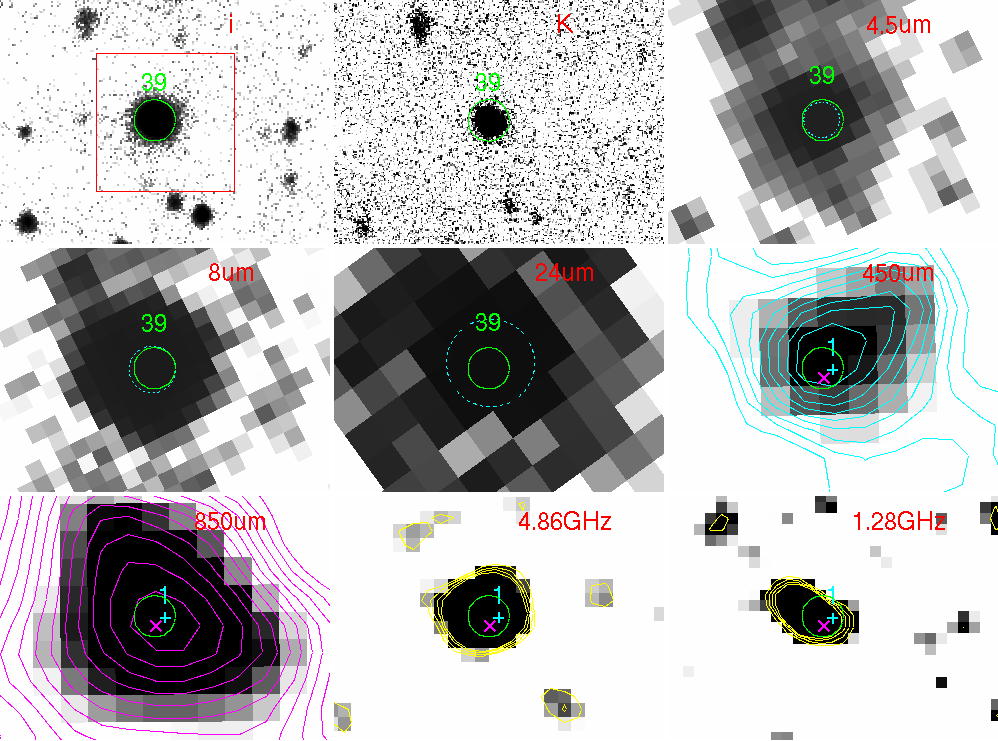}
\caption{Finding charts for 850\_1/450\_1. From top-left to
bottom-right: $i$ (the size of the red box is $10''\times10''$), $K$,
4.5\um{} (dashed cyan circle is the positions of the 4.5\um{} sources
with radius equal to the flux extraction radius), 8\um{} (dashed cyan
circles mark 8\um{} sources and extraction radii), 24\um{} (dashed
cyan circles mark 24\um{} sources and extraction radii), 450\um{}
(with cyan contours starting at 2$\sigma$ and increasing in steps of
0.5), 850\um{} (with magenta contours, starting at 2$\sigma$ and
increasing in steps of 0.5), 4.86~GHz (with yellow contours starting
at 2$\sigma$ and in intervals of 1), 1.29~GHz (with yellow contours,
same as previous). All finding charts have the same FOV and the solid
circles indicate the optical/NIR source position and extraction radius
(green corresponds to 1.5$''$, red to smaller radii), only the sources
mentioned in the text have been labelled. The labels have been omitted
in the SCUBA and radio images for clarity, but we have labelled and
marked instead with cyan ``+'' our positions for the 450\um{} and with
magenta ``x" the positions of the 850\um{} sources.}\label{Figfc450-1}
\end{figure*}

\begin{figure*}
  \includegraphics[width=18cm]{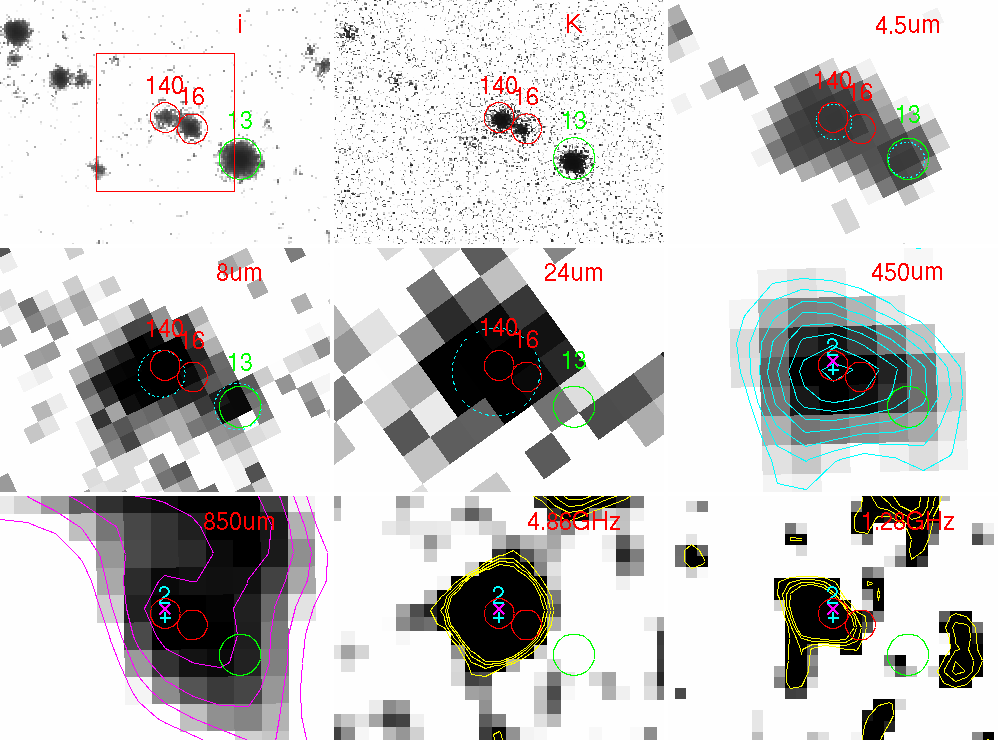}
\caption{Finding charts for 850\_2/450\_2. Same layout as in Fig. \protect{\ref{Figfc450-1}}.}\label{Figfc450-2}
\end{figure*}

\begin{figure*}
  \includegraphics[width=18cm]{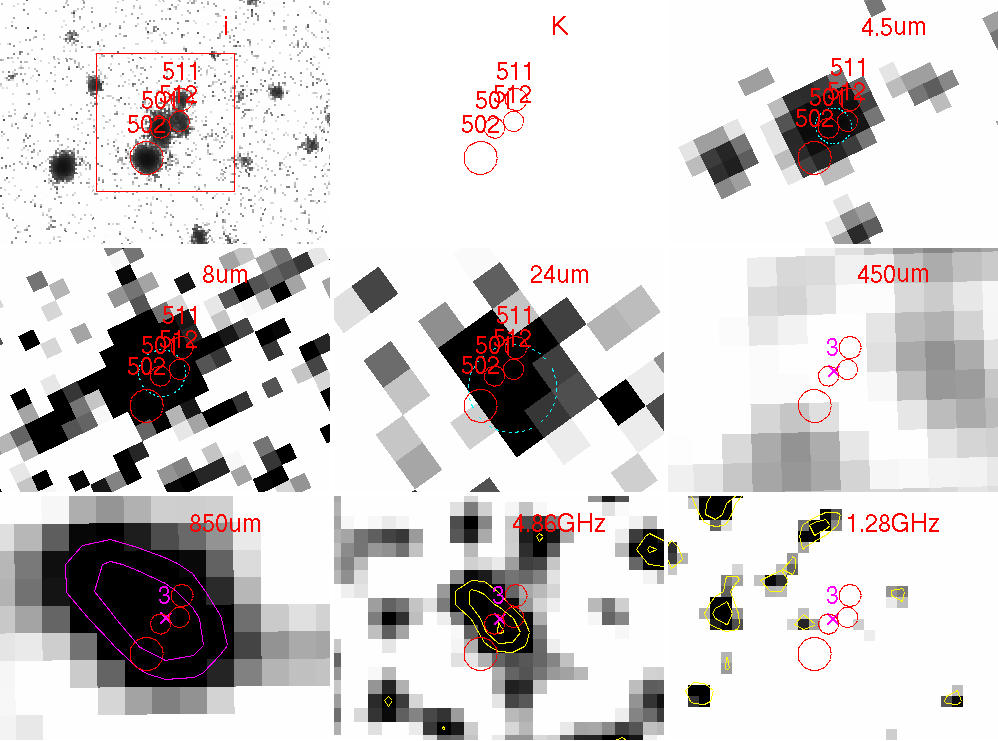}
\caption{Finding charts for 850\_3. Same layout as in Fig. \ref{Figfc450-1}.}   \label{Figfc850-3}
\end{figure*}

\begin{figure*}
  \includegraphics[width=18cm]{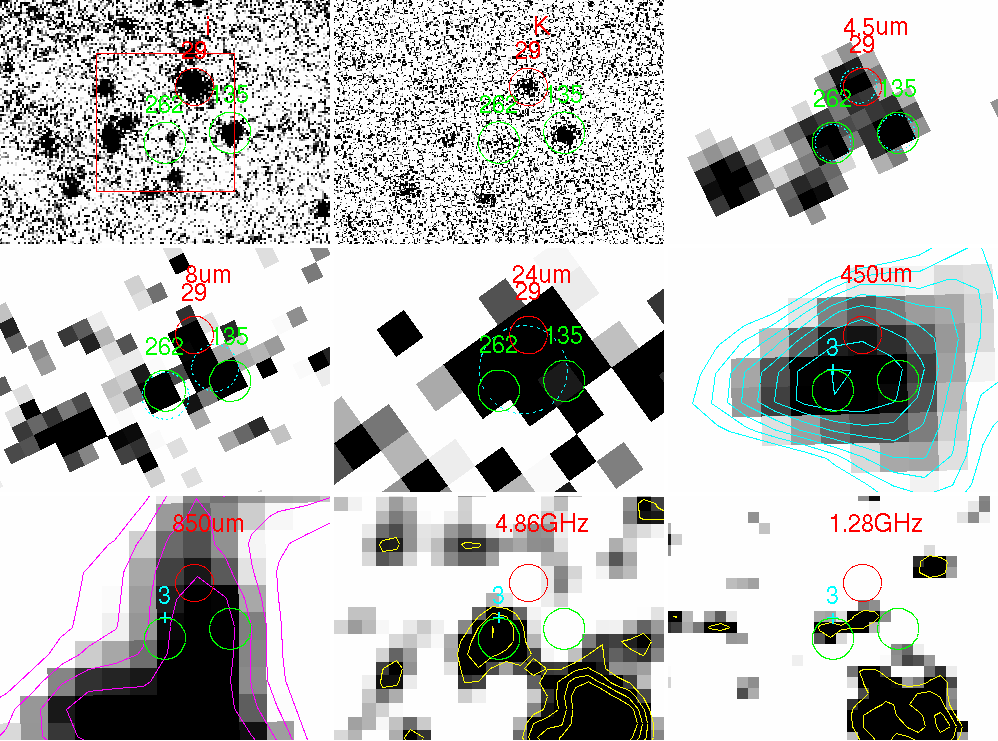}
\caption{Finding charts for 450\_3. Same layout as in Fig. \ref{Figfc450-1}.}\label{Figfc450-3}
\end{figure*}

\begin{figure*}
  \includegraphics[width=18cm]{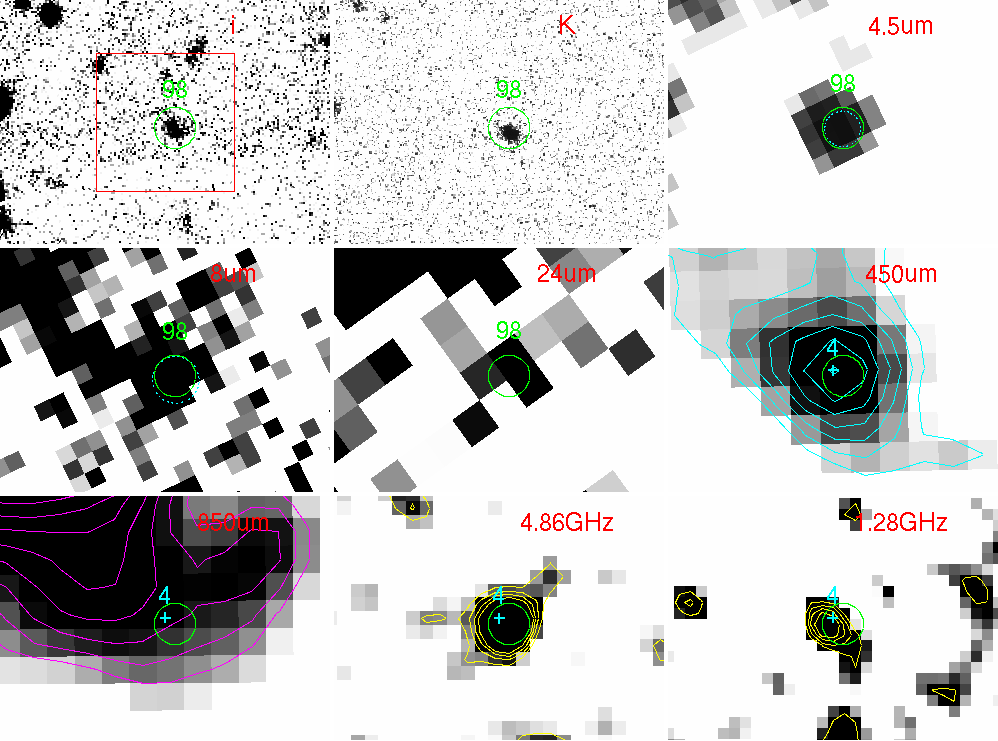}
\caption{Finding charts for 450\_4. Same layout as in Fig. \ref{Figfc450-1}.}  \label{Figfc450-4}
\end{figure*}

\subsection{Association to submm sources, SEDs and SFR}
\label{SEDs}

We have plotted finding charts in $i$, $K$, \Spitzer{}, SCUBA, VLA and
GMRT for the 5 distinct sources from S10 (see Figures \ref{Figfc450-1}
to \ref{Figfc450-4}). The best fit (to $RiZJK$, 4.5 and 8\um{})
photometric redshift templates are shown as red lines in
Fig. \ref{FigSED}, as well as the observed wide-band SEDs with a
number of templates.  All the redshift confidence intervals given in
the rest of this Section are $1\sigma$.

The templates from MRR08 do not include dust emission, so in order to
model adequately the longer wavelengths we have used the starburst
models of \citet{CE01} (henceforth CE01), which have been obtained
from observations at several NIR-FIR wavelengths requiring the
relative fluxes in different bands to be consistent with the observed
IR luminosity functions. One feature of these models which is worth
noting is that their shape is luminosity-dependent and hence, given a
redshift and the flux in one band, the SED is fully defined. The
authors have provided {\sc IDL} routines to compute their templates at
a given redshift and through a number of instrumental bandpasses. We
have added the SCUBA 450 and 850\um{} sensitivity curves to the set of
available bandpasses. The absence of a good correlation between
the B-band and IR luminosities implies that the optical/NIR part of
their SEDs is uncertain. This is not a major problem here since we are
only using the CE01 SEDs for the longer wavelengths. Instead of
performing a formal $\chi^2$ fit to the data, we have computed the
CE01 templates corresponding to each of the observed \Spitzer{} and SCUBA
fluxes at the redshift of each source. 

In the case of 850\_1/450\_1 we expect to have a strong AGN component,
specially in the rest-frame optical/NIR part of the spectrum. We have
modelled it using templates from MRR08. We have approximated their
RR1v2 AGN template with a broken power-law (which is a good approximation
to the continuum shape), redshifting it to $z=1.82$ (the redshift of
RX~J0941). Within the Unified Model for AGN \citep{Anton93} part of
this direct emission is intercepted by a dusty torus, which
``reprocess'' it and emits it at mid and far-infrared wavelengths. To
model this emission we have used the dust torus template from MRR08.

The SFR from the CE01 templates normalized to 450/850\um{} are given
in Table~\ref{TabSCUBA}. The SFR have been calculated using the
following expression from \citet{K98}

\begin{equation}
SFR(M_\odot/\mathrm{ y})=1.7217\times 10^{-10}L_\mathrm{ FIR}(L_\odot),
\end{equation}

\noindent where $L({\mathrm 8-1000\mu m})$ ($L_\mathrm{ IR}$) and
$L({\mathrm 40-500\mu m})$ ($L_\mathrm{ FIR}$) have been estimated
from the 12, 24, 60 and 100\um{} fluxes (obtained from the CE10
templates normalized to the SCUBA fluxes) using the formula described
in \citet{SM96}  with $C=1.4$. Most of our submm sources would be
classified as Hyper Luminous IR Galaxies using the criteria in
\citet{MRR00}. Since the CE01 templates are simply rescaled to the
observed SCUBA fluxes ($F$), the uncertainties can be estimated via
$\Delta$SFR/SFR=$\Delta F/F$.

We now discuss the counterparts to the submm
sources and the SEDs individually:

\begin{itemize}

\item{850\_1/450\_1:} The sharp radio contours prove that this source
is indeed the central QSO (RX~J0941). It is associated with source 39
(242/180 in S10).  The full SED for this object is shown in
Fig.~\ref{FigSED} (top left panel).  The dashed red line
represents the CE01 template normalized to the observed 450\um{}
flux at $z=1.82$. It clearly reproduces well the observed emision at
450\um{} and redwards, but it falls well short of the observed values
bluewards of that wavelength. We have fitted the $RiZJK$ fluxes with
our broken power-law model for the ``direct'' AGN emission from
MRR08 (green dashed line), which more or less follow the blue
spectrum in the observed optical/NIR part of the SED, but again fails
to account for the observed MIR emission. We have added the starburst
emission to the direct AGN emission, subtracted it from the 8\um{} point
and rescaled the dust torus emission from MRR08 to this value
(blue dashed line). The sum of these three components is shown as the
black solid line in the Figure. It clearly follows the overall shape
of the SED, which cannot be reproduced by any two of the models
alone. The submm emission is clearly dominated by the starburst, with
the total AGN contribution being at least two orders of magnitude
smaller, as originally stated in \citet{P01}. There is very little
room for an additional radio contribution from the AGN in this
radio-quiet QSO.

\item{850\_2/450\_2:} The radio data show a strong pointlike source,
coincident with our source 140 (191/155 in S10). Its photometric
redshift is $z_p=1.85$ with a redshift interval spanning 1.83-2.16
with spectral type Sab. Our source 16 is about 2.2~arcsec away from
140, and could be contributing to its 24~$\mu$m flux. Its photometric
redshift is 3.0 (2.9-3.0, Scd). Source 13 is brighter and coincides
with elongations in the contours at \450um{} and \850um. Its
photometric redshift is $z_p\sim1.19\pm0.03$ (Scd), and hence is
probably unrelated to the central QSO.  In Fig.~\ref{FigSED} (top
right panel) the dashed and dotted lines show the CE01 templates
normalized to 450 and 850\um{} (respectively). These lines straddle
very closely all bluer and redder wavelengths.

\item{850\_3:} The NE-SW-elongated 2 and 2.5$\sigma$ \850um{} contours
are aligned with the VLA 2$\sigma$ contour. However, there is no sign
of it in the GMRT image. This peak would favour our source 501 as the
prefered counterpart ($z_p=1.85$, $z$ in 1.83-2.20, Sab, 339/241 in
S10), although this optical/NIR neighborhood is rather crowded:
sources 502, 511 and 512 are also close. The \Spitzer{} emission
appears pointlike favouring either 501 or 512. Unfortunately, the
absence of $J$ and $K$ data for this area does not allow to put strong
constraints on the photometric redshift range for 512 ($z_p\sim0.7$,
SB). In Fig.~\ref{FigSED} (middle left panel) the CE01 template
normalized to the 850\um{} flux (blue dotted line) reproduces
reasonably well all spectral points redwards of 8\um, but falls well
short at bluer wavelengths, although the spectral shape is
similar. Normalizing the CE01 template to the 4.5\um{} flux is a bit
too high at 24\um{} and 850\um, and too high at radio wavelengths.

\item{450\_3:} The 450 and \850um{} contours are complex (this source
might also be detected in \850um, see Section \ref{data}) The
$>2\sigma$ VLA contour includes the \450um{} peak of 450\_3,
coinciding with our source 262 (206/160 in S10), and hence is our
prefered counterpart.  Object 262 comes originally from a 4.5\um{}
detection and only has upper limits in $iZJ$, hence its redshift is
poorly constrained, seeming to prefer $z_p\sim2.8$ (youngE,
$z>2.6$). Its absolute $B$ magnitude would be too faint for both
$z=1.82$ and $z=2.8$, suggesting an even higher redshift. This source
seems to be connected by a $2\sigma$ bridge of radio emission to the
radio ``blob" discussed in Section~\ref{RadioResults}.  Nearby sources
with some evidence for \Spitzer{} emission are 135 (205/159 in S10,
$z_p=1.85$,1.8-2.2, Scd) and 29 (to the NW and along a N-S elongation
in the \850um contours, $z_p=2.7$, 2.6-2.9, SB). Object 135 coincides
with the main E-W asymmetry of the \450um{} emission and could be
related to the central QSO, although the absence of a radio
detection does not support this idea.  Since the redshift of 262 is
poorly constrained we show the CE01 SED normalized to the IRAC
4.5~$\mu$m and SCUBA points both at $z=2.78$ (bottom left) and at the
redshift of the QSO $z=1.82$ (bottom right). The CE01 template
normalized to the 450\um{} point is well above the radio and the 4.5
and 8\um{} points in both cases, while when it is normalized to the
4.5\um{} point it matches better the IRAC and the radio points. This
may mean that there might be some contribution to the SCUBA flux from
our source 135, in agreement with the elongated contours at
450\um. The 24\um{} source position is intermediate between 262, 135
and 29, so the flux in that range also probably includes contribution
from them, explaining why it lies above the CE01 template at
$z=2.78$.

\item{450\_4:} Coincides with a well defined pointlike IRAC, VLA and
GMRT source, very close to our source
98 (the unlabeled source to the W of the centre of the finding chart
for 450\_4 in S10). We have found $z_p=1.85$
(1.8-2.1, Sbc). Its SED is plotted in Fig.~\ref{FigSED} (middle right panel), showing  
 that the CE01 models at that
redshift neatly follow the shape of the \Spitzer{} and SCUBA points
(only the one normalized to the SCUBA point is shown, the models
normalized to the \Spitzer{} points are indistinguishable in this
scale).  The only clearly discrepant range
is radio, which falls a factor of about 3 below the CE01
template.
\end{itemize}

In summary, the radio positions allow assigning unambiguous
counterparts to all of the submm sources. Four out of five submm
sources (850\_1/450\_1 -RX~J0941-, 850\_2/450\_2, 450\_4 and 850\_3)
have counterparts with redshifts compatible with 1.82 within $\sim
1\sigma$. The three latter ones have very similar spectral shapes,
compatible with an existing relatively old stellar population with
ongoing star formation. The last submm source (450\_3) appears to be a
radio emitting object at higher redshift, although there is a nearby
counterparts within the ``right" redshift interval and very similar
properties to the other counterparts to the submm sources (135). From
the number counts in \citet{Coppin06} we would expect to have about
one submm source unrelated to the QSO in our field, so it is not
surprising that 450\_3 turns out to be a ``background" object,
considering also its radio faintness.

\subsection{Upper limits on AGN luminosities and estimating black hole masses}
\label{AGNX-ray}

\begin{figure}
  \includegraphics[width=9cm]{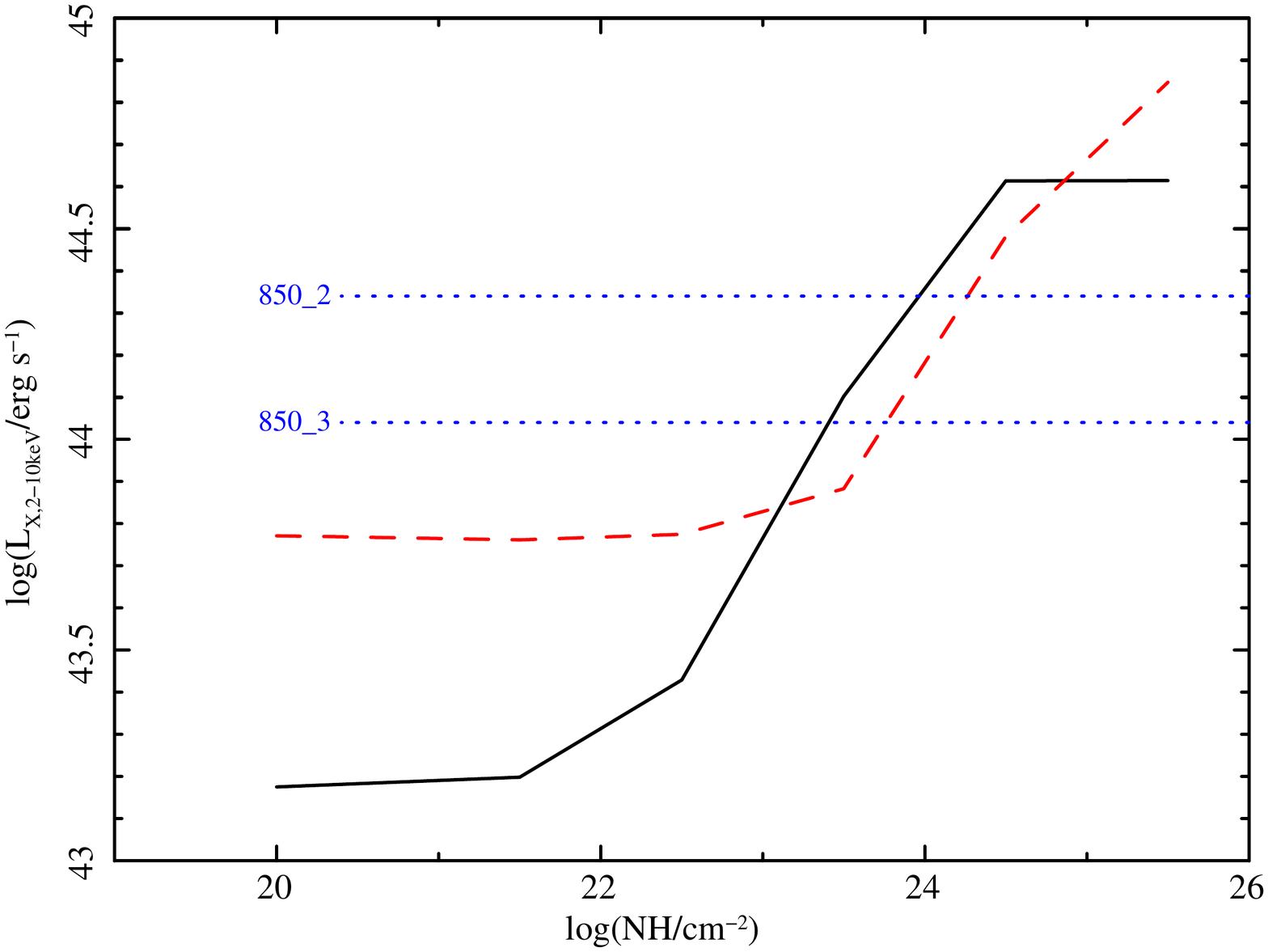}

  \caption{Upper limits to the 2-10~keV intrinsic luminosity of an AGN
  from the upper limits to the soft band (black solid line) and hard band (red
  dashed line) count rates from {\sl XMM-Newton observations} versus
  the intrinsic column density. The value for $N_H=0$ has been plotted
  at $\log(N_H/{\mathrm cm}^{-2})=20$. The blue dotted lines mark the limits from the
  observed 24\um{} fluxes for sources 850\_2 and 850\_3, see text for an explanation.} \label{LumNH}
  \end{figure}

850\_1/450\_1 is the central QSO and hence it obviously harbours
a powerful AGN. In this Section we will concentrate on the rest of our
submm sources.  The ``blob" of radio emission close to 450\_3 might
be interpreted as evidence for the presence of an AGN in this
object. This possibility cannot be rejected using our current data due
to its quality at the bluer wavelengths, where the AGN would be
expected to dominate.

The optical-to-radio SEDs of the other three counterparts to the SCUBA
sources have very limited room left for an additional AGN emission
component. We will derive below limits to the luminosity of any
putative AGN and the mass of the associated SMBH, under the assumption
that the latter is growing through accretion and radiation is emitted
in the process. Our galaxies could harbour SMBH more massive than
deduced below if they were dormant.

We have taken advantage of our {\sl XMM-Newton} observation of this
field \citep{P10} to set limits on the intrinsic luminosity of any
such AGN (see Section \ref{data}). Of course, the central QSO is
actually detected \citep[see][]{P10}. Three other SCUBA sources have
upper limits on the pn camera (850\_3 is in a gap between two chips in
that camera), with similar values of $<$0.00034 ($<$0.000455) cts/s in
the 0.5-2~keV -soft- (2-10~keV -hard-) band (using a simple power-law
with photon index $\Gamma=1.9$ and local Galactic absorption, these
values correspond to fluxes of about $<6\times10^{-16}$ and
$<4\times10^{-15}$~erg~cm$^{-2}$~s$^{-1}$, respectively). We have
taken these countrates and estimated what would be the intrinsic
luminosity of a $z=1.82$ AGN to produce those countrates if viewed
through different amounts of intrinsic absorption, parameterized as
the Hydrogen column density $N_H$. We have used $N_H$=0, and
$\log(N_H/\mathrm{ cm}^{-2})=21.5-25.5$ in steps of 1~dex.  We adopt a
simply parametrised AGN spectral model following \citet{Gilli07},
i.e. an exponentially cut-off absorbed power law of photon index
$\Gamma=1.9$, a small scattered component, a reflection component, and
an Fe K emission line. We include a correction for Compton scattering
of the continuum at the highest column densities.  We have used the
program {\sc xspec} and an on-axis response matrix and ancillary file
for pn to convert from countrates in those bands to luminosities in
the rest-frame 2-10~keV band $L_\mathrm{ X,2-10keV}$. The results are
shown in Fig.~\ref{LumNH}.

Supernovae and binary stars associated with star formation produce
X-rays, which would further reduce the allowance for any AGN
emission. \citet{Ranalli03} found a tight correlation between X-ray
luminosity and FIR luminosity in local starburst galaxies which, using
Eq. 1 above \citep{K98}, can be expressed as SFR$(M_\odot/\mathrm{
y})=2\times10^{-40}L_\mathrm{ X,2-10keV}$(erg/s) (their equation
15). For the typical values of our galaxies of SFR$\sim 2000M_\odot$/yr
this corresponds to $L_\mathrm{ X,2-10keV}\sim10^{43}$~erg/s,
comparable with the above upper limits for the soft band and moderate
absorption.


In addition, we can set very conservative upper limits to the presence
of an obscured AGN from the observed 24\um{} flux. We have used as
fiducial values those of sources 850\_2 and 850\_3 ($\sim$700 and
$\sim350\mu$Jy, respectively), because 450\_4 is not detected in that
band and 450\_3 is probably a background source and its emission in
that band is probably a blend of several sources. Assuming that the
emission from dust heated by the AGN at 24\um{} is isotropic and that
the full observed fluxes at that wavelength have this origin, we can
use an AGN SED (the \citealt{Richards06} RQ QSO template redshifted to
$z=1.82$) to find out what hard X-ray fluxes those 24\um{} fluxes
would correspond to. The corresponding values are
$\log(L_\mathrm{X,2-10keV}/\mathrm{erg\,s^{-1}})=44.3$ and 44.0,
respectively. Since the starburst SED template leaves very limited
room for an additional AGN contribution (for instance, for 850\_2 the
difference between the 24~$\mu$m flux of CE01 template normalized to
450~$\mu$m and the actual observed flux at that wavelength is about
200~$\mu$Jy, a factor of about two below our lowest fiducial value),
this is an absolute ceiling to the maximum luminosity of a putative
AGN in the innermost regions of our submm sources.

We conclude that AGN with intrinsic 2-10~keV luminosity
$\sim10^{43}$~erg/s could be present in our SCUBA sources, if they are
behind modest amounts of absorption ($\log(N_H/\mathrm{ cm}^{-2})\leq
22-23$), while objects with almost QSO luminosities
($\sim10^{44}$~erg/s) would have to be hiding behind much more
substantial amounts of material ($\log(N_H/\mathrm{
cm}^{-2})>24$). More luminous AGN are not compatible with our current
data. An X-ray luminosity of $\sim10^{44}$~erg/s is compatible with
the observed hard X-ray luminosities of SMGs at similar X-ray fluxes
\citep{Alexander05a}

We now translate these limits on X-ray luminosity into a limit on the
putative SMBH mass. The first step is to estimate the total bolometric
luminosity of the AGN, which we can write as $L_\mathrm{
AGN}=\kappa L_\mathrm{ X,2-10keV}$, where $\kappa$ is the bolometric
correction. The maximum theoretical luminosity emitted by material falling into a
black hole of mass $M_\bullet$ is called Eddington luminosity
$L_{Edd}$ \citep{Eddington13,Rees84} and can be expressed as

\begin{equation}
\log\left({L_\mathrm{ Edd}\over L_\odot}\right)=4.53+\log\left({M_\bullet \over M_\odot}\right).
\end{equation}

\noindent Of course, a source can be emitting below this limit. The
Eddington ratio $\eta$ is defined as $\eta=L_{AGN}/L_{Edd}$, and hence

\begin{equation}
\log\left({M_\bullet \over M_\odot}\right)=\log\left({L_\mathrm{ X,2-10keV}\over L_\odot}\right)-4.53+\log(\kappa/\eta).
\label{EqBHmassLX}
\end{equation}

\noindent Replacing the upper limit found above ($L_X<10^{44}$~erg/s$=10^{10.42}\,L_\odot$) we finally get

\begin{equation}
\log\left({M_\bullet \over M_\odot}\right)<5.89+\log(\kappa/\eta).
\end{equation}

\citet{Alexander08} study the black hole masses, Eddington ratios and
X-ray luminosities of obscured SMGs at $z\sim2$, finding that for
objects with $L_X\sim10^{44}$~erg/s, $\log(M_\bullet/M_\odot)\sim 7.8$
with $\eta\sim0.2$. They also find similar values of $\eta$ for nearby
obscured ULIRGs. Finally, the gas-rich environment in which the SMBH
are likely to reside and the large fraction of SMGs that host AGN
activity (see \citealt{Alexander05a,B05}, but see also
\citealt{Laird10}) suggest that the accretion is likely to be
reasonably efficient, so we adopt a fiducial value of $\eta=0.2$.
Using the luminosity-dependent relation between X-ray
luminosity and bolometric luminosity of \citet{Marconi04}, gives
$\kappa=33$, very similar to the value of $\kappa=35$ which would be
deduced from the QSO SED of \citet{Elvis94}, as used by \citet{Alexander08}. \citet{VF07}
find $\kappa=15-25$ for AGN emitting at low Eddington ratio
($\eta\la0.1$) and $\kappa=40-70$ for higher Eddington ratios. We will assume $\kappa=35$, getting

\begin{equation}
\log\left({M_\bullet \over M_\odot}\right)<8.1+\log\left({\kappa\over 35}\right)-\log\left({\eta \over 0.2}\right).
\label{EqBHmass}
\end{equation}

\noindent Our logarithmic upper limit on the SMBH masses can be as
high as 8.4 ($\eta=0.2,\kappa=70$) or as low as 7.4
($\eta=1,\kappa=35$) for efficient accretion. For inefficient
accretion (assuming $\kappa=15$) we get 8.0 for $\eta=0.1$ and 9.0 for
$\eta=0.01$.

\subsection{Dust masses}
\label{Mdust}

We have estimated dust masses from the observed SCUBA fluxes using Eq. 5 in \citet{MS09}:

\begin{equation}
M_\mathrm{ dust}={ d_L^2(z) F_\nu(\nu) \over (1+z) \kappa(\nu(1+z)) B_\nu(T,\nu(1+z))},
\end{equation}

\noindent where

\begin{equation}
\kappa(\nu(1+z))=\kappa_0\times\left({\nu(1+z) \over \nu_0}\right)^\beta,
\end{equation}

\noindent and $B_\nu(T,\nu)$ is the usual blackbody emissivity
\citep[e.g. Eq. 3 in][]{MS09}.

For $F_\nu$ we have used the monochromatic fluxes at $\lambda$=450 and
850\um{} from Table~\ref{TabSCUBA}.  In order to estimate a range of
dust masses for each object, we have used the eight possible
combinations of ($\kappa_0=0.04$~m$^2$/kg, $\lambda_0=$1.2~mm)
\citep{Beelen06} and ($\kappa_0=2.64$~m$^2$/kg, $\lambda_0=$125\um)
\citep{Dunne03} with ($T$=47K, $\beta=1.6$) \citep[representative of
unobscured quasars at high redshift][]{Beelen06} and ($T$=35K,
$\beta$=1.5) (representative SMGs; see
\citealt{K06}) and with the minimum and the maximum of the
$1\sigma$ uncertainty interval in their photometric redshifts, taking
the minimum and the maximum values of those combinations (which always
corresponded to the minimum redshift, higher $\kappa_0$, higher $T$ and
maximum redshift, higher $\kappa_0$, lower $T$, respectively). We have
made two exceptions: 850\_1/450\_1 (the QSO) for which we have fixed
$z=1.82$, and 450\_3 for which we give both the intervals fixing
$z=1.82$ (main body of the table) and allowing the full photometric
redshift uncertainty interval (footnote).

An additional source of uncertainty comes from the errors on the submm
fluxes, which range from about 10\% to about 40\% (with one case
reaching 70\%). These fractional errors are generally smaller than the
uncertainty ranges derived above, and therefore we take the latter to
represent a good estimate of the order of magnitude of the dust masses
present in our sources.

For the sources detected at both submm wavelengths, we find that the
dust mass estimates derived from 450 and 850\um{} are compatible within
the uncertainties. All sources have dust masses in the range
$2\times 10^8-3\times 10^9$ solar masses.

We can also estimate the gas mass present in the central regions of
those galaxies from the dust masses, assuming a gas-to-dust ratio of
54, deduced by \citet{K06} for $z=1-3$ SMG (with an uncertainty of
about 20\%). This value is similar to the one obtained by
\citet{Seaquist04} for the central regions of nearby SCUBA
galaxies. For our typical dust mass of $\sim10^9M_\odot$ this
corresponds to a gas mass of $\sim5\times10^{10}M_\odot$.

\subsection{Galaxy stellar masses}
\label{Mstar}

We have estimated stellar masses from the absolute $K$-band magnitudes
$M_K$ using the expression

\begin{equation}
\log\left({M_*\over M_\odot}\right)={M_K-3.28 \over -2.5}-\log(3.2),
\end{equation}

\noindent which we have taken from \citet{B05} assuming a
light-to-mass ratio $L_K/M=3.2\times L_{K,\odot}/M_\odot$, which is
intermediate between the average values from their fits to a burst and
a continuous star-forming model.

We have obtained $M_K$ at the best photometric redshift integrating
the best fit template using {\sc hyperz}. These estimates should be
reasonably accurate for the sources with IRAC detections up to
$z\sim 2.6$, since the best fit template would then be ``anchored" in a
photometric point on a neighbouring rest frame wavelength.

No estimate of the uncertainty in the normalization of the best fit
template is returned by {\sc hyperz}. Our first approximation to the
uncertainty in our derived stellar mass comes from the uncertainty in
the redshift:

\begin{equation}
\Delta\log(M_*)={\Delta M_K \over 2.5}\sim 2\Delta\log(d_L(z)),
\end{equation}

\noindent since $M_K=m_K-5\log(d_L(z))$ (distance modulus). We have
estimated the uncertainty in the luminosity distance as half the
difference between the luminosity distances at the lower and upper
limits of the $1\sigma$ redshift interval.

Since essentially the derived stellar mass is proportional to the flux $F$
at the observed wavelength corresponding to the rest-frame $K$-band
(between 4.5 and 8\um{} at $z=1.82$), a second contribution
to its uncertainty would be:

\begin{equation}
\Delta\log(M_*)'={\Delta\log F}={1\over \ln10}{\Delta F \over F}.
\end{equation}

\noindent This corresponds to about 0.03, 0.04 and 0.07 for 450\_4, 850\_2
and 850\_3, for total errors of 0.09, 0.10 and 0.13, respectively.

A further source of uncertainty is the assumed light-to-mass
ratio. MRR08 use a different approach to estimate the stellar masses
from their best fit templates 3.6\um{} luminosities, with a
light-to-mass ratio that depends on the galaxy class and on the cosmic
time. We have also estimated the stellar masses of our 3 counterparts
following this recipe. We obtain $L/M=0.05-0.06$ and
$\log\left({M_*/M_\odot}\right)=11.85,12.06,12$ for 450\_4, 850\_2
and 850\_3,
respectively. These stellar masses are between 2 and 3 times larger than those
in Table~\ref{TabSCUBA}, the discrepancies being larger than the
total error budget calculated above.

In what follows, we will use for our sources a fiducial value
$\log(M_*/M_\odot)=11.5\pm0.2$ which takes into account the different
stellar mass values for the different sources and the redshift and
photometric uncertainties discussed above. Using the MRR08 recipe
would produce values higher by 0.3-0.5~dex.

These stellar masses are similar to those obtained for other samples
of SMGs: e.g., \citet{B05} find $\log(M_*/M_\odot)=11.4\pm0.4$ in the
CDF-N/GOODS-N region, and \citet{MHW10} find $\log(M_*/M_\odot)=11-12$
with a median of 11.7, despite using very different SEDs and methods
for deriving the stellar mass (in the second case). As pointed out by
\citet{B05} those stellar masses are about 10 times larger than those
of typical UV-selected star-forming galaxies at similar redshifts
\citep{Shapley05}. They are compatible with Schechter stellar mass for
the GOODS-MUSIC galaxies at $z\sim1.8$
\citep[$\log(M_*/M_\odot)\sim11.3\pm0.15$][]{Fon06}. They also similar
to the Schechter mass of 3.6-4.5\um-selected galaxies in the HDF-N,
CDF-S and the Lockman Hole in the $1.6<z\leq2.0$ redshift interval
\citep[$\log(M_*/M_\odot)=11.40\pm0.18$][]{PPG08}. Their $0<z<0.2$
galaxies have slightly smaller masses
($\log(M_*/M_\odot)=11.16\pm0.25$), although compatible within
$1\sigma$.

Although the above masses are stellar masses for the whole galaxy,
they are probably dominated by the spheroid \citep[as discussed
by][]{B05,Alexander08}, since high-angular resolution
$H$-band imaging of SMGs suggests that the stellar structure of most SMGs
is best described by a spheroid/elliptical galaxy light distribution
\citep{Swinbank10}.

\section{Evolutionary stage of the counterparts}
\label{Discussion}

In this section we use our estimates of the SFR and the stellar and
dust mass for the counterparts to the submm sources to investigate
their current state, their plausible future development and their
relation to any putative AGN.
 
An immediate conclusion is that our submm galaxies are very mature;
they have stellar masses typical of MIR and submm-selected galaxies
($\sim 3\times10^{11}M_\odot$), and their gas masses
($\sim5\times10^{10}M_\odot$) are around an order of magnitude
smaller than our estimates of the stellar mass, indicating that
$\ga$90\% of the maximum possible stellar mass is already in place.

Interestingly, \citet{Onodera10} find a galaxy at a very similar
redshift ($z=1.82$) with a very similar stellar mass, which has
properties fully consistent with those expected for passively evolving
progenitors of present day giant ellipticals. When the vigorous star
formation taking place in our SMGs is over, they might
well look like that object.

A common indicator of the strength of the star formation in galaxies
is the Specific Star Formation Rate (SSFR) defined as the ratio of the
SFR to the stellar mass. It is generally found that high mass galaxies
tend to have lower SSFR at $z\la2$ \citep[see e.g.][]{Feulner05,PPG08,Ilbert10},
which is interpreted as a sign that star formation does not
significantly change their stellar mass over this redshift
interval. Our SMGs have SSFR$\sim3-10$Gyr$^{-1}$. 
These values are about an order of magnitude higher
than those of MIR-selected galaxies with similar redshifts
($1.6<z\leq2$) and stellar masses ($\log(M_*/M_\odot)>11$) in
\cite{PPG08}. Our SMGs are amongst the most actively star forming
objects of their time.

\citet{Ilbert10} find that, for a fixed range in SSFR, the fraction of
high mass objects among their starforming galaxies drops with redshift
between $z=2$ and $z=1$, while that of lower mass objects is
approximately constant down to $z\sim1$. This implies that the low
mass star-forming galaxies are able to maintain a high SSFR, while the
massive galaxies evolve rapidly into systems with a lower SSFR (since
the mass cannot go down, this means necessarily lower SFR). This is
the expected evolution of our SMGs as deduced above from an
independent argument, and follows a clear ``downsizing" pattern
\citep[e.g.][]{Cowie96,PPG08}

Assuming that our SMGs host a SMBH which is growing and emiting
radiation, in Section \ref{AGNX-ray} (equation \ref{EqBHmass}) we
deduced a very conservative upper limit to the mass of the SMBH in AGN
in the centre of our galaxies of $\log(M_\bullet/M_\odot)<8.1$. Hence,
their black hole mass-to-galaxy mass ratio is
$\log(M_\bullet/M_*)<-3.4$. Applying the relation between those
quantities found by \citet{MH04} for local galaxies we get
$\log(M_\bullet/M_*)=-2.6$ (for $\log(M_*/M_\odot)=11.5$), so at face
value our sources seem to have black hole masses at least about a
factor of 6 below local galaxies of the same mass. Other studies have
also found lower SMBH-to-galaxy mass values in $z\sim2$ SMG
\citep[e.g.][]{B05,Alexander08,Coppin09}, which are commonly
interpreted as the black hole growth lagging behind the galaxy growth,
since galaxies appear to be essentially fully grown while AGN are
"under size".

There are considerable uncertainties in such estimates \citep[as
discussed at some length by e.g.][]{B05,Alexander08}. We include
0.2~dex of uncertainty in the stellar masses of our SMGs and the
upper limit to the SMBH mass ranges between about 7.4 and 8.4 (in
logarithmic solar masses) for different Eddington ratios and
bolometric corrections (see Sections \ref{AGNX-ray} and \ref{Mstar}).
Taking the extreme values for both masses, we get upper limits to
$\log(M_\bullet/M_*)$ between -2.9 and -4.3, while the value for local
galaxies is $\log(M_\bullet/M_*)_\mathrm{ local}=-2.6\pm0.3$ (using
the rms intrinsic dispersion quoted in \citealt{MH04}, using instead
\citealt{HR04} we would get $-2.5\pm0.3$).

Hence the upper limit to the ratio we derived above between the local
and our SMBH-to-galaxy mass is $\log(M_\bullet/M_*)_\mathrm{
local}/\log(M_\bullet/M_*)>6$, with an attached ``uncertainty
interval" $\sim 1-100$. The AGN in the centres of our galaxies would
have to be heavily obscured and just at the limit of detection for
that ratio to be close to unity. The X-ray obscured AGN in SMGs
studied by \citet{Alexander05a} are indeed heavily obscured (80\% have
$\log(N_H/$cm$^{-2})>23$) but their average luminosity is $L_\mathrm{
X,2-10keV}\sim5\times10^{43}$~erg/s, leading to an additional factor
of 2 on that ratio. Finally, another factor of 2 or 3 would be
necessary if we used the MRR08 method for estimating the stellar
masses.

In summary, our SMGs appear to be mature
$\log(M_*/M_\odot)=11.5\pm0.2$ and with limited scope of significant
further increase in mass (remaining gas mass
$\sim5\times10^{10}M_\odot$), while any putative AGN in their centres
($\log(M_\bullet/M_\odot)\la8.1$) is probably smaller than those of
local galaxies of similar mass (by a factor of more than 6) but with
plenty of fuel to grow if it can tap the infered gas mass: the mass
that a SMBH at our derived upper limit needs to accrete to reach the local
SMBH-to-gas mass ratio is only about 1.3\% of the total gas mass.

The QSO RX~J0941 shows strong X-ray emission (luminosity of $3\times
10^{44}$~erg/s), high SFR (a few thousand solar masses per year) and
strong ionized winds \citep{P10}. Within the evolutionary scheme
outlined in the Introduction, all these clues would mean that the SMBH
is starting to push away the material responsible for the star
formation and the SMBH growth.

Finally, the properties of our SMGs are similar to those of similar
objects in ``blank field" observations \citep[e.g.][ whose properties
have been obtained with similar methods]{B05,Alexander08} despite
being in a significant overdensity. Similar properties of
``spike'' and ``field'' SMGs at $z\sim1.99$ have also been reported by
e.g. \citet{Chapman09}.  This indicates that the environment does not
affect substantially the SMG properties. Galaxy interactions are
commonly invoked to channel material to the internal regions of
galaxies, triggering starbursts \citep{MH94}. We have mixed evidence:
RX~J0941 (850\_1) and 450\_4 appear isolated in the optical/NIR bands,
while 850\_2 and 850\_3 are in crowded immediate environments.
Unfortunately, we lack high angular resolution data around our sources
to be able to constraint tightly their interaction history.

\section{Conclusions}
\label{Conclusions}

Using X-ray to radio data, we have studied  the immediate environment of
RX~J0941, a $z=1.82$ QSO with strong ionized winds visible in the UV
and in X-rays \citep{P10}, which shows as well both strong submm
emission and an overdensity of submm galaxies (SMGs) around it
\citep{S04,S10}. Such high density regions are expected to form a
galaxy cluster at $z=0$ \citep{K96}.

The 6 and 20cm radio data confirm the presence of all the submm
sources and help to pinpoint their observed-frame optical-to-MIR
counterparts. We have obtained photometric estimates of their
redshifts using {\sc hyperz} to fit SWIRE galaxy templates \citep{MRR08}
to $RiZJK$ and \Spitzer{}-IRAC data.

We found that four of the five unique submm sources are indeed
associated with the QSO. The fifth source appears to be a background
source, perhaps with an associated lobe of radio emission. The
photometric redshifts appear to be robust against using different sets
of templates \citep{BC03,MRR08}.

Under the assumption that a growing Super Massive Black Hole -SMBH- is
hosted by our SMGs, we have used X-ray upper limits and the observed
24\um{} fluxes to estimate a very conservative upper limit to the
X-ray luminosity ($<10^{44}$erg/s) of such putative AGN (and hence a
SMBH mass of $\log(M_\bullet/M_\odot)<8.1$), even if Compton thick
material is absorbing the direct AGN emission.

We have used the rest-frame optical-to-radio starburst templates of
\citet{CE01} to estimate the star formation rates (SFR) of the host
galaxy of the QSO and the SMGs, obtaining individual rates of
1000$M_\odot$/yr or above (see Table \ref{TabSCUBA}). We have also estimated dust
($\sim10^9M_\odot$) and gas ($\sim5\times10^{10}M_\odot$) masses,
assuming parameters typical of the centres of SMGs.

We have also estimated their stellar masses from their rest-frame
$K$-band luminosities, finding $\log(M_*/M_\odot)\sim 11.5\pm0.2$, in
line with other mass estimates for SMGs \citep[e.g.][]{B05}, but about
10 times larger than typical UV-selected star-forming galaxies at
similar redshifts.

An immediate conclusion (as also found for other samples of SMGs) is
that they are mature galaxies, with large stellar masses and an
inferred reservoir of gas that would only allow a further 10\%
increase in mass at most.

At the same time, their SMBH masses are a least factor of $\sim 6$
below that expected for their galaxy mass and the local SMBH-to-galaxy
mass ratio \citep{MH04}. This has also been found previously for other
samples of SMGs, and is commonly interpreted as the growth of the SMBH
lagging behind the galaxy growth, since the latter is already fully
matured while the AGN still requires substantial growth
\citep{B05,Alexander08,Coppin09}. Accretion of a few percent of the
infered gas mass on to the SMBH would be sufficient to reach the local
mass ratio.

Further observations are required to understand better the issues
discussed here. Deeper X-ray data would allow to place tighter limits
on, or to actually detect, AGN emission from the SMGs.
Forthcoming Herschel SPIRE and PACS data will ``fill in" the gaps in the
FIR-submm SED, getting better estimates of the SFR and dust
masses. High angular resolution millimetre observations would give
more accurate estimates of the gas mass. Photometric redshifts and SED
studies for the other overdensities of SMGs in our sample \citep{S10}
will assess their relationship with the central QSOs and their
evolutionary status, helping to understand the role of these high
density peaks in the formation and evolution of galaxies.

\section*{Acknowledgments}

The authors thank the anonymous referee for helpful suggestions to
clarify the paper.  F.J.C. thanks M. Rowan-Robinson for his help with
the SWIRE templates, R. Pell\'o for her help with {\sc hyperz} and
N. Ben\'\i{}tez for his help with {\sc bpz}.  F.J.C., J.E. and
S.F. acknowledge financial support from the Spanish Ministerio de
Educaci\'on y Ciencia (later Ministerio de Ciencia e Innovaci\'on)
under projects ESP2006-13608-C02-01 and AYA2009-08059. F.J.C., J.A.S.,
and M.J.P. acknowledge further support from the Royal Society. Based
on observations made at the William Herschel Telescope and at the
Isaac Newton Telescope which is operated on the island of La Palma by
the Isaac Newton Group in the Spanish Observatorio del Roque de los
Muchachos of the Instituto de Astrof\'\i{}sica de Canarias. Also based
on observations obtained at the Gemini Observatory (under program
GN2004A-Q-52), which is operated by the Association of Universities
for Research in Astronomy, Inc., under a cooperative agreement with
the NSF on behalf of the Gemini partnership: the National Science
Foundation (United States), the Science and Technology Facilities
Council (United Kingdom), the National Research Council (Canada),
CONICYT (Chile), the Australian Research Council (Australia),
Minist\'{e}rio da Ci\^{e}ncia e Tecnologia (Brazil) and Ministerio de
Ciencia, Tecnolog\'{i}a e Innovaci\'{o}n Productiva (Argentina). UKIRT
is operated by the Joint Astronomy Centre, Hilo, Hawaii on behalf of
the UK Science and Technology Facilities Council. Also based on
observations made with the Spitzer Space Telescope, which is operated
by the Jet Propulsion Laboratory, California Institute of Technology,
under NASA contract 1407. The James Clerk Maxwell Telescope is
operated by The Joint Astronomy Centre on behalf of the Science and
Technology Facilities Council of the United Kingdom, the Netherlands
Organisation for Scientific Research, and the National Research
Council of Canada. JCMT data were taken under project ID M03AU46. Also
based on data collected at the XMM-Newton, an ESA science mission with
instruments and contributions directly funded by ESA Member States and
NASA. This research has made use of data obtained from the SuperCOSMOS
Science Archive, prepared and hosted by the Wide Field Astronomy Unit,
Institute for Astronomy, University of Edinburgh, which is funded by
the UK Science and Technology Facilities Council. The NRAO is a
facility of the National Science Foundation operated under cooperative
agreement by Associated Universities. We thank the staff of the GMRT
who have made these observations possible. GMRT is run by the National
Centre for Radio Astrophysics of the Tata Institute of Fundamental
Research. This publication makes use of data products from the Two
Micron All Sky Survey, which is a joint project of the University of
Massachusetts and the Infrared Processing and Analysis
Center/California Institute of Technology, funded by the National
Aeronautics and Space Administration and the National Science
Foundation. Funding for the SDSS and SDSS-II has been provided by the
Alfred P. Sloan Foundation, the Participating Institutions, the
National Science Foundation, the U.S. Department of Energy, the
National Aeronautics and Space Administration, the Japanese
Monbukagakusho, the Max Planck Society, and the Higher Education
Funding Council for England. The SDSS is managed by the Astrophysical
Research Consortium for the Participating Institutions. The
Participating Institutions are the American Museum of Natural History,
Astrophysical Institute Potsdam, University of Basel, University of
Cambridge, Case Western Reserve University, University of Chicago,
Drexel University, Fermilab, the Institute for Advanced Study, the
Japan Participation Group, Johns Hopkins University, the Joint
Institute for Nuclear Astrophysics, the Kavli Institute for Particle
Astrophysics and Cosmology, the Korean Scientist Group, the Chinese
Academy of Sciences (LAMOST), Los Alamos National Laboratory, the
Max-Planck-Institute for Astronomy (MPIA), the Max-Planck-Institute
for Astrophysics (MPA), New Mexico State University, Ohio State
University, University of Pittsburgh, University of Portsmouth,
Princeton University, the United States Naval Observatory, and the
University of Washington.

\appendix

\section[]{Aperture matching of the photometric points}
\label{AppMatchPhot}

Since each of our 5 optical-NIR images has a different seeing, we need
to smooth all of them so that the effective seeing is the same as the
one with the worst value ($Z$ in our case). We have done this using
{\sc gaus} in {\sc iraf}. 

We have chosen a default aperture of 1.5$''$ radius, which corresponds
to $\sim$98\% enclosed flux for a 2D gaussian with FWHM equal to the
effective seeing (or 96\% for the empirical aperture correction
explained below). There are 75 sources closer than 3$''$ to another
source, so that the default aperture would assign counts from the same
pixels to at least two sources. To avoid this, for each source we have
chosen an aperture which is the minimum of 1.5$''$ and half the distance
to the closest source (in steps of 0.1$''$). The minimum aperture found using this method is
0.6$''$. 


The flux $F$ of each source is determined (using {\sc phot} within
{\sc iraf}) from the number of counts within the chosen aperture
(after background subtraction), and the magnitude from the usual
$-2.5\log F$. The default {\sc phot} procedure to obtain the
uncertainty in the magnitude involves determining the mean value
(msky) and the standard deviation (stdev) of the pixel values over the
background area. This value can also be used to estimate the 1$\sigma$
limit for the flux of each source as $\sqrt{\mathrm{ msky \times
area}}$ (assuming Poisson statistics) where area is the area for the
extraction of the source counts.

Following this recipe produces both too low values for the magnitude
errors and unrealistically low flux limits. We believe that this is
because the reduction processes outlined in Section \ref{data} imply
co-adding many individual images with re-scalings to match median or
average pixel values over essentially arbitrary areas. Furthermore,
since we have smoothed the images to have the same ``effective''
seeing, we have introduced correlations between the values of
neighbouring pixels, and essentially averaging away the dispersion in
the sky values.  This is bound to alter the statistical
characteristics of the counts in each final pixel, with respect to the
na\"ive Poisson assumption underlying most photometry packages (like
{\sc phot}).

We have followed an alternative approach. For each image we have
excluded an area of radius 2.5$''$ around each source and, for each
aperture radius, we have placed at random a large number (700) of
circles of that size on the image, and we have measured the standard
deviation (skydev) of the values of the fluxes within the simulated
apertures, scaled to the total circle area. We have only kept
simulated apertures in which more than 99\% of their area was outside
the excluded zones. The number of simulated apertures has been chosen
not to oversample the un-excluded area. We have checked that reducing
the fraction of good pixels to 90\%, increasing the exclusion radius
to 3$''$ or increasing the number of simulated apertures to 2000 does
not change the obtained values. The error on the fluxes can then be
estimated as $\sqrt{F/\mathrm{ epadu+skydev^2}}$ and the 1$\sigma$
upper limits as $m_0-2.5\log_{10}(\mathrm{ skydev})$, where epadu is
the e$^-$-per-ADU of the detector and $m_0$ are the zero point
magnitudes given in Table~\ref{TableData}. The 1$\sigma$ limits
obtained in this way are shown in Table~\ref{TableUpLims} in terms of
magnitudes. In order to get, for example, 5$\sigma$ limits,
$2.5\log_{10}(5)\sim1.75$ needs to be subtracted from those values.

This problem is also present in the \Spitzer{} images, so we have
obtained $skydev$ for them in the same way as for the optical-NIR
images, using the source extraction radii given in Section
\ref{photcal}. Since those images were in units of MJy/sr, the errors
on the fluxes obtained by {\sc SExtractor} (assuming Poisson
statistics) were underestimated, and very similar for all sources
(since they were dominated by the variance of the pixel values in the
background). We have used the {\tt FLUXCONV} and {\tt GAIN} keywords
in the headers of the 4.5 and 8~$\mu$m images\footnote{{\tt
http://ssc.spitzer.caltech.edu/irac/iracinstrumenthandbook/home/}} and
the exposure times to estimate the original number of counts in the
image, and the corresponding errors. The total uncertainty in the
image fluxes was obtained by adding to these errors in quadrature the
total uncertainty in the determination of the values of those keywords
(about 5\%, see footnote) and the uncertainty in the sky values
(skydev). These fluxes and errors are the ones used for the
empirical flux calibration described in \ref{photcal}. After applying
this calibration, the 1$\sigma$ sky standard deviations in the 4.5, 8
and 24\um{} bands are 0.5, 2.1 and 52~$\mu$Jy, respectively.

In order to obtain a good estimate of the stellar mass it is essential
to have a good measurement of the rest-frame $K$-band absolute
magnitude, which would fall between the IRAC bands
at the redshift of RX~J0941. Hence it is essential to include those
bands in the photometric redshift fits for the 101 sources detected in
either of those two MIR bands. But, since the  \Spitzer{} fluxes are
corrected to ``infinite" aperture, good estimates of this correction
for the optical-NIR fluxes are needed as well, so that all the
photometric points fed to the fit match. 

We have done this empirically, extracting for each source a radial
profile in the smoothed $i$-band image between $0.2''$ and half the
distance to the nearest source (see above) and fitting to it both a
single and a double gaussian profile, allowing thus for both ``nuclear"
and ``galaxy" components. We have chosen for each source the best fit
between those two using an F-test.  The average value of the $\sigma$
of a single gaussian profile (for the sources for which this was the
best fit) is $0.59''$, slightly larger than the one used for matching
the angular resolution of all optical-NIR images ($0.54''$), but
essentially compatible with it.  A correction factor was then calculated
from the best fit profile for each source.

The fluxes/magnitudes of the 101 sources with 4.5 or 8\um{}
detections have been corrected with this factor before the photometric
redshift fitting, so the derived physical magnitudes are already for
"infinite" aperture. For the remaining 138 sources, this factor has
been applied to these derived magnitudes after the photometric
redshift fit.

\begin{table}
 \centering
 \begin{minipage}{80mm}
  \caption{1$\sigma$ upper limits for each image and aperture (see
App. \ref{AppMatchPhot})}
  \begin{tabular}{rrrrrr}
  \hline
Aperture & $R$ & $i$ & $Z$ & $J$ & $K$ \\ 
(arcsec)\\
 \hline
0.6 & 27.06 & 27.41 & 26.26 & 24.67 & 23.44\\
0.7 & 26.76 & 27.05 & 26.02 & 24.43 & 23.08\\
0.8 & 26.55 & 26.86 & 25.85 & 24.13 & 22.86\\
0.9 & 26.28 & 26.73 & 25.64 & 23.93 & 22.65\\
1.0 & 26.15 & 26.49 & 25.49 & 23.74 & 22.51\\
1.1 & 25.95 & 26.38 & 25.40 & 23.60 & 22.34\\
1.2 & 25.80 & 26.25 & 25.21 & 23.47 & 22.21\\
1.3 & 25.68 & 26.14 & 25.10 & 23.32 & 22.08\\
1.4 & 25.51 & 25.99 & 25.01 & 23.17 & 21.99\\
1.5 & 25.40 & 25.89 & 24.89 & 23.06 & 21.91\\
 \hline
\end{tabular}
\end{minipage}
\label{TableUpLims}
\end{table}


\bsp

\label{lastpage}

\end{document}